\documentclass[journal,onecolumn,draftcls,12pt]{IEEEtran}

\usepackage{xspace}
\usepackage{url}
\usepackage[cmex10]{amsmath}
\usepackage{bbm}
\usepackage{graphicx}
\usepackage{epstopdf}
\usepackage{paralist}
\usepackage[normalem]{ulem}
\usepackage{fancyref} 
\usepackage{amsmath}
\usepackage{amssymb}
\usepackage[acronym]{glossaries}
\usepackage[stretch=16,shrink=16,step=4]{microtype}

\usepackage{color}
\definecolor{links}{rgb}{0.7,0,0}   
\definecolor{urls}{rgb}{0,0,0.8}    
\definecolor{cites}{rgb}{0,0,0.8}   
\usepackage[colorlinks,hyperindex,linkcolor=links,citecolor=cites,urlcolor=urls]{hyperref} 

\usepackage{dsfont}
\usepackage{footnote}
\usepackage{color}
\usepackage{balance}
\usepackage{float}
\usepackage{cite}
\usepackage{wasysym}     
\usepackage{amssymb}

\usepackage{vmr-symbols-vecbold}
\usepackage{standard-macros}
\usepackage{mathbbol}

\usepackage{tikz}
\usetikzlibrary{patterns}
\usetikzlibrary{shapes,arrows,positioning}
\usetikzlibrary{calc}
\usetikzlibrary{fit}
\usetikzlibrary{plotmarks}
\tikzset{every picture/.style={font issue=\scriptsize, >=stealth},font issue/.style={execute at begin picture={#1\selectfont}}}
\tikzset{three sided left/.style={
        draw=none,
        xshift=\pgflinewidth,
        append after command={
            [shorten <= -0.5\pgflinewidth]
            ([shift={(-1.5\pgflinewidth,-0.5\pgflinewidth)}]\tikzlastnode.north east) edge ([shift={( 0.5\pgflinewidth,-0.5\pgflinewidth)}]\tikzlastnode.north west) 
            ([shift={( 0.5\pgflinewidth,-0.5\pgflinewidth)}]\tikzlastnode.north west) edge ([shift={( 0.5\pgflinewidth,+0.5\pgflinewidth)}]\tikzlastnode.south west)            
            ([shift={( 0.5\pgflinewidth,+0.5\pgflinewidth)}]\tikzlastnode.south west) edge ([shift={(-1.0\pgflinewidth,+0.5\pgflinewidth)}]\tikzlastnode.south east)
        }}}
        
\tikzset{three sided right/.style={
        draw=none,
        xshift=-\pgflinewidth,
        append after command={
            [shorten <= -0.5\pgflinewidth]
            ([shift={( 1.5\pgflinewidth,-0.5\pgflinewidth)}]\tikzlastnode.north west) edge ([shift={(-0.5\pgflinewidth,-0.5\pgflinewidth)}]\tikzlastnode.north east) 
            ([shift={(-0.5\pgflinewidth,-0.5\pgflinewidth)}]\tikzlastnode.north east) edge ([shift={(-0.5\pgflinewidth,+0.5\pgflinewidth)}]\tikzlastnode.south east)            
            ([shift={(-0.5\pgflinewidth,+0.5\pgflinewidth)}]\tikzlastnode.south east) edge ([shift={( 1.0\pgflinewidth,+0.5\pgflinewidth)}]\tikzlastnode.south west)
        }}}

\usepackage{pgfplots}
\usepgfplotslibrary{fillbetween}
\pgfplotsset{
  compat=newest, 
  width=\columnwidth,    
  height=0.8\columnwidth,   
  plot coordinates/math parser=false,
  standard/.style={
    axis equal,
    axis line style=help lines,
    axis x line=center,
    axis y line=center,
    axis z line=center},
    grid style={dashed,gray},
    minor grid style={dotted,gray},
    major grid style={dotted,gray},
    ylabel absolute, ylabel style={yshift=-0.4cm},
    xlabel absolute, xlabel style={yshift=0.25cm}
}

\makeatletter
 \pgfdeclarepatternformonly[\tikz@pattern@color,\LineSpace,\LineWidth]{my horizontal lines}%
    {\pgfpointorigin}{\pgfqpoint{100pt}{1pt}}{\pgfqpoint{100pt}{\LineSpace}}%
    {
        \pgfsetcolor{\tikz@pattern@color}
        \pgfsetlinewidth{\LineWidth}
        \pgfpathmoveto{\pgfqpoint{0pt}{0.5pt}}
        \pgfpathlineto{\pgfqpoint{100pt}{0.5pt}}
        \pgfusepath{stroke}
    }
 
 \pgfdeclarepatternformonly[\tikz@pattern@color,\LineSpace,\LineWidth]{my vertical lines}%
    {\pgfpointorigin}{\pgfqpoint{1pt}{100pt}}{\pgfqpoint{\LineSpace}{100pt}}%
    {
        \pgfsetcolor{\tikz@pattern@color}
        \pgfsetlinewidth{\LineWidth}
        \pgfpathmoveto{\pgfqpoint{0.5pt}{0pt}}
        \pgfpathlineto{\pgfqpoint{0.5pt}{100pt}}
        \pgfusepath{stroke}
    } 
 
 \pgfdeclarepatternformonly[\tikz@pattern@color,\LineSpace,\LineWidth]{my grid}%
    {\pgfqpoint{-1pt}{-1pt}}{\pgfqpoint{\LineSpace}{\LineSpace}}
    {\pgfqpoint{\LineSpace}{\LineSpace}}%
    {
        \pgfsetcolor{\tikz@pattern@color}
        \pgfsetlinewidth{\LineWidth}
        \pgfpathmoveto{\pgfqpoint{0pt}{0pt}}
        \pgfpathlineto{\pgfqpoint{0pt}{\LineSpace + 0.1pt}}
        \pgfpathmoveto{\pgfqpoint{0pt}{0pt}}
        \pgfpathlineto{\pgfqpoint{\LineSpace + 0.1pt}{0pt}}
        \pgfusepath{stroke}
    }
 
 \pgfdeclarepatternformonly[\tikz@pattern@color,\LineSpace,\LineWidth]{my north east lines}
    {\pgfqpoint{-\LineWidth}{-\LineWidth}}{\pgfqpoint{\LineSpace}{\LineSpace}}
    {\pgfqpoint{\LineSpace}{\LineSpace}}%
    {
        \pgfsetcolor{\tikz@pattern@color}
        \pgfsetlinewidth{\LineWidth}
        \pgfpathmoveto{\pgfqpoint{-\LineWidth}{-\LineWidth}}
        \pgfpathlineto{\pgfqpoint{\LineSpace + 0.1pt}{\LineSpace + 0.1pt}}
        \pgfusepath{stroke}
    }
 
 \pgfdeclarepatternformonly[\tikz@pattern@color,\LineSpace,\LineWidth]{my north west lines}
    {\pgfqpoint{-\LineWidth}{-\LineWidth}}{\pgfqpoint{\LineSpace}{\LineSpace}}
    {\pgfqpoint{\LineSpace}{\LineSpace}}%
    {
        \pgfsetcolor{\tikz@pattern@color}
        \pgfsetlinewidth{\LineWidth}
        \pgfpathmoveto{\pgfqpoint{-\LineWidth}{\LineSpace}}
        \pgfpathlineto{\pgfqpoint{\LineSpace + 0.1pt}{-\LineWidth}}
        \pgfusepath{stroke}
    }
 
 \pgfdeclarepatternformonly[\tikz@pattern@color,\LineSpace,\LineWidth]{my crosshatch}%
    {\pgfqpoint{-1pt}{-1pt}}{\pgfqpoint{\LineSpace}{\LineSpace}}
    {\pgfqpoint{\LineSpace}{\LineSpace}}%
    {
        \pgfsetcolor{\tikz@pattern@color}
        \pgfsetlinewidth{\LineWidth}
        \pgfpathmoveto{\pgfqpoint{\LineSpace + 0.1pt}{0pt}}
        \pgfpathlineto{\pgfqpoint{0pt}{\LineSpace + 0.1pt}}
        \pgfpathmoveto{\pgfqpoint{0pt}{0pt}}
        \pgfpathlineto{\pgfqpoint{\LineSpace + 0.1pt}{\LineSpace + 0.1pt}}
        \pgfusepath{stroke}
    }
 
 \pgfdeclarepatternformonly[\tikz@pattern@color,\LineSpace,\PointSize]{my dots}%
    {\pgfqpoint{-\LineSpace*0.25}{-\LineSpace*0.25}}
    {\pgfqpoint{\LineSpace*0.25}{\LineSpace*0.25}}
    {\pgfqpoint{\LineSpace*0.75}{\LineSpace*0.75}}%
    {
        \pgfsetcolor{\tikz@pattern@color}
        \pgfpathcircle{\pgfqpoint{0pt}{0pt}}{\PointSize}
        \pgfusepath{fill}
    }
\makeatother
 
\newdimen\LineSpace
\newdimen\PointSize
\newdimen\LineWidth
\tikzset{
    line space/.code={\LineSpace=#1},
    line space=3pt
}
\tikzset{
    point size/.code={\PointSize=#1},
    point size=.5pt
}
\tikzset{
    pattern line width/.code={\LineWidth=#1},
    pattern line width=.4pt
}

\DeclareSymbolFontAlphabet{\amsmathbb}{AMSb}%

\newcommand{\lro}[1]{\lefto({#1}\right)}																
\newcommand{\lrbo}[1]{\lefto \lbrace {#1} \right \rbrace}															
\newcommand{\lrho}[1]{\lefto [ {#1} \right ]}																				

\newcommand{\lr}[1]{\left({#1}\right)}																
\newcommand{\lrb}[1]{\left \lbrace {#1} \right \rbrace}															
\newcommand{\lrh}[1]{\left [ {#1} \right ]}																				

\safemath{\dopplerspread}{B_D}																								
\safemath{\delayspread}{T_D}																									
\safemath{\nc}{n\sub{c}}																										
\safemath{\nd}{n\sub{d}}																										
\safemath{\ntx}{n\sub{t}} 																											
\safemath{\nrx}{n\sub{r}}																											
\safemath{\ntxt}{\tilde{n\sub{t}}}																											
\safemath{\cb}{\ensuremath{L}} 																								
\safemath{\cl}{\ensuremath{n}} 																								
\safemath{\txanto}{{\ensuremath{\tilde{m}_t}}} 																		
\safemath{\cs}{M} 																														
\safemath{\idPustm}{\ensuremath{S_{k}}}
\safemath{\error}{\ensuremath{\epsilon}} 																				
\safemath{\eexp}{\ensuremath{\mathcal{E}}} 																			
\safemath{\nsubc}{n\sub{s}}			 																						
\safemath{\nofdm}{n\sub{o}} 																									
\safemath{\bc}{\ensuremath{B_c}} 																							
\safemath{\ts}{\ensuremath{T_s}} 																							
\safemath{\nrb}{\ensuremath{n_{rb}}} 																						
\safemath{\nres}{\ell}
\newcommand{\cgauss}[2]{\mathcal{CN}\lro{\ensuremath{#1, #2}  }}   								
\safemath{\maxk}{M^*\lr{\nres, \nsubc, \nofdm, \epsilon, \rho}}
\safemath{\Rmax}{R^*}
\safemath{\Emin}{E\sub{b}^*/N_0}
\safemath{\np}{\ensuremath{n\sub{p}}}
\safemath{\code}{\ensuremath{\mathcal{C}}}
\safemath{\err}{\ensuremath{\epsilon}}
\safemath{\rp}{\ensuremath{\rho\sub{p}}}
\safemath{\rd}{\ensuremath{\rho\sub{d}}}
\safemath{\mI}{\ensuremath{i\lro{\randvecy ; \randvecx}}} 				

\safemath{\randveca}{\bm{A}}
\safemath{\randvecb}{\bm{B}}
\safemath{\randvecc}{\bm{C}}
\safemath{\randvecd}{\bm{D}}
\safemath{\randvece}{\bm{E}}
\safemath{\randvecf}{\bm{F}}
\safemath{\randvecg}{\bm{G}}
\safemath{\randvech}{\bm{H}}
\safemath{\randveci}{\bm{I}}
\safemath{\randvecj}{\bm{J}}
\safemath{\randveck}{\bm{K}}
\safemath{\randvecl}{\bm{L}}
\safemath{\randvecm}{\bm{M}}
\safemath{\randvecn}{\bm{N}}
\safemath{\randveco}{\bm{O}}
\safemath{\randvecp}{\bm{P}}
\safemath{\randvecq}{\bm{Q}}
\safemath{\randvecr}{\bm{R}}
\safemath{\randvecs}{\bm{S}}
\safemath{\randvect}{\bm{T}}
\safemath{\randvecu}{\bm{U}}
\safemath{\randvecv}{\bm{V}}
\safemath{\randvecw}{\bm{W}}
\safemath{\randvecx}{\bm{X}}
\safemath{\randvecy}{\bm{Y}}
\safemath{\randvecz}{\bm{Z}}
\safemath{\randvecphi}{\bm{\Phi}}

\safemath{\randmatA}{\amsmathbb{A}}
\safemath{\randmatB}{\amsmathbb{B}}
\safemath{\randmatC}{\amsmathbb{C}}
\safemath{\randmatD}{\amsmathbb{D}}
\safemath{\randmatE}{\amsmathbb{E}}
\safemath{\randmatF}{\amsmathbb{F}}
\safemath{\randmatG}{\amsmathbb{G}}
\safemath{\randmatH}{\amsmathbb{H}}
\safemath{\randmatI}{\amsmathbb{I}}
\safemath{\randmatJ}{\amsmathbb{J}}
\safemath{\randmatK}{\amsmathbb{K}}
\safemath{\randmatL}{\amsmathbb{L}}
\safemath{\randmatM}{\amsmathbb{M}}
\safemath{\randmatN}{\amsmathbb{N}}
\safemath{\randmatO}{\amsmathbb{O}}
\safemath{\randmatP}{\amsmathbb{P}}
\safemath{\randmatQ}{\amsmathbb{Q}}
\safemath{\randmatR}{\amsmathbb{R}}
\safemath{\randmatS}{\amsmathbb{S}}
\safemath{\randmatT}{\amsmathbb{T}}
\safemath{\randmatU}{\amsmathbb{U}}
\safemath{\randmatV}{\amsmathbb{V}}
\safemath{\randmatW}{\amsmathbb{W}}
\safemath{\randmatX}{\amsmathbb{X}}
\safemath{\randmatY}{\amsmathbb{Y}}
\safemath{\randmatZ}{\amsmathbb{Z}}
\safemath{\randmatSigma}{\mathbb{\Sigma}}
\safemath{\randmatPhi}{\mathbb{\Phi}}
\safemath{\randmatLambda}{\mathbb{\Lambda}}

\safemath{\matSigma}{\bm{\Sigma}}
\safemath{\matPhi}{\bm{\Phi}}
\safemath{\matLambda}{\bm{\Lambda}}

\def\bie{\begin{IEEEeqnarray}{rCl}}
\def\eie{\end{IEEEeqnarray}}

\usepackage{pgfplots}
\usepgfplotslibrary{groupplots}
\usetikzlibrary{pgfplots.groupplots}

\usepgfplotslibrary{external}
  
\newcommand\marksize{1} 
\newcommand\linew{1pt} 
\newcommand{\fwidth}{\textwidth}

\makeatletter
\def\@IEEEinterspaceratioM{0.265}
\def\@IEEEinterspaceMINratioM{0.1651}
\def\@IEEEinterspaceMAXratioM{0.38}

\def\@IEEEinterspaceratioB{0.31}
\def\@IEEEinterspaceMINratioB{0.19}
\def\@IEEEinterspaceMAXratioB{0.38}
\@IEEEtunefonts
\makeatother
\let\abs\undefined
\newcommand{\abs}[1]{\lvert#1\rvert}		

\makeglossaries

\newacronym{lte}{LTE}{{long term evolution}}
\newacronym{ue}{UE}{user equipment}
\newacronym{ul}{DL}{uplink}
\newacronym{dl}{DL}{downlink}
\newacronym{3gpp}{3GPP}{{3rd generation partnership project}}
\newacronym{rb}{RB}{resource block}
\newacronym{tti}{TTI}{transmission time interval}
\newacronym{ofdm}{OFDM}{orthogonal frequency-division multiplexing}
\newacronym{iid}{i.i.d.}{identical and independently distributed}
\newacronym{psd}{PSD}{power spectral density}
\newacronym{mimo}{MIMO}{multiple-input multiple-output}
\newacronym{bler}{BLER}{block error probability}
\newacronym{mtc}{MTC}{machine-type communication}
\newacronym{csi}{CSI}{channel state information}
\newacronym{dvbt}{DVB-T}{{terrestrial digital video broadcasting}}
\newacronym{pat}{PAT}{pilot-assisted transmission}
\newacronym{ml}{ML}{{maximum likelihood}}
\newacronym{dt}{DT}{{dependence-testing}}
\newacronym{ustm}{USTM}{{unitary space-time modulation}}
\newacronym{siso}{SISO}{{single-input single-output}}
\newacronym{nn}{SNN}{{scaled nearest-neighbor}}
\newacronym{rce}{RCEE}{{random-coding error exponent}}
\newacronym{grce}{GRCEE}{{generalized random-coding error exponent}}
\newacronym{rcus}{RCUs}{random-coding union bound with parameter $s$}
\newacronym{rcu}{RCU}{{random-coding union bound}}
\newacronym{osd}{OSD}{ordered statistics decoding}
\newacronym{llr}{LLR}{log-likelihood ratio}
\newacronym{qpsk}{QPSK}{quaternary phase shift keying}
\newacronym{nlos}{NLOS}{non-line-of-sight}
\newacronym{los}{LOS}{line-of-sight}
\newacronym{cgf}{CGF}{cumulant-generating function}
\newacronym{gmi}{GMI}{generalized mutual information}
\newacronym{pdf}{pdf}{probability density function}

\makeatletter
\let\MYcaption\@makecaption
\makeatother

\usepackage[font=scriptsize]{subcaption}

\makeatletter
\let\@makecaption\MYcaption
\makeatother
\let\marginnote\undefined

\newcommand{\marginnote}[1]{\marginpar{\framebox{\framebox{#1}}}}

\begin{document}
\IEEEoverridecommandlockouts
\title{Short Packets over Block-Memoryless Fading Channels: Pilot-Assisted or Noncoherent Transmission?}
\author{Johan \"Ostman,~\IEEEmembership{Student Member,~IEEE}, Giuseppe~Durisi,~\IEEEmembership{Senior Member,~IEEE}, Erik~G.~Str\"om,~\IEEEmembership{Senior Member,~IEEE}, Mustafa C. Co\c{s}kun,~\IEEEmembership{Student Member,~IEEE}, and  Gianluigi~Liva,~\IEEEmembership{Senior Member,~IEEE}
\thanks{This work was partly supported by the Swedish Research Council under grants 2014-6066 and 2016-03293.}
\thanks{The material of this paper was presented in part at the IEEE International Workshop on Signal Processing Advances in Wireless Communications, July 2017, Sapporo, Japan~\cite{Ostman17-03a}.}

\thanks{Johan \"Ostman, Giuseppe Durisi, and Erik G. Str{\"o}m are with the Department of Electrical Engineering, Chalmers University of Technology, Gothenburg 41296, Sweden (e-mail: \{johanos,durisi,erik.strom\}@chalmers.se).}
\thanks{Mustafa C. Co\c{s}kun and Gianluigi Liva are with the Institute of Communications and Navigation of the German Aerospace Center (DLR), M{\"u}nchner Strasse 20, 82234 We{\ss}ling, Germany (e-mail: mustafa.coskun@tum.de, gianluigi.liva@dlr.de).}
}
\maketitle

\begin{abstract}

We present nonasymptotic upper and lower bounds on the maximum coding rate achievable when transmitting short packets over a Rician memoryless block-fading channel for a given requirement on the packet error probability.
We focus on the practically relevant scenario in which there is no \emph{a priori} channel state information available at the transmitter and at the receiver.
An upper bound built upon the min-max converse is compared to two lower bounds: the first one relies on a noncoherent transmission strategy in which the fading channel is not estimated explicitly at the receiver; the second one employs pilot-assisted transmission (PAT) followed by maximum-likelihood channel estimation and scaled mismatched nearest-neighbor decoding at the receiver.
Our bounds are tight enough to unveil the optimum number of diversity branches that a packet should span so that the energy per bit required to achieve a target packet error probability is minimized, for a given constraint on the code rate and the packet size.
Furthermore, the bounds reveal that noncoherent transmission is more energy efficient than PAT, even when the number of pilot symbols and their power is optimized. 
For example, for the case when a coded packet of $168$ symbols is transmitted using a channel code of rate $0.48$ bits/channel use, over a  block-fading channel with block size equal to $8$ symbols,  PAT requires 
an additional $1.2$ dB of energy per information bit to achieve a packet error probability of $10^{-3}$ compared to a suitably designed noncoherent transmission scheme.
Finally, we devise a PAT scheme based on punctured tail-biting quasi-cyclic codes and ordered statistics decoding, whose performance are close ($1\dB$ gap at $10^{-3}$ packet error probability) to the ones predicted by our PAT lower bound. 
This shows that the PAT lower bound provides useful guidelines on the design of actual PAT schemes.

\end{abstract}

\section{Introduction} 
\label{sec:introduction}
Supporting the transmission of short packets under stringent latency and reliability constraints is critically required for next-generation wireless communication networks to address the needs of future autonomous systems, such as connected vehicles, automated factories and smart grids~\cite{metis-project-deliverable-d1.113-04a, durisi16-09a}. 
%
%
%
%
%
Classic information-theoretic performance metrics, i.e., the  \emph{ergodic} and the \emph{outage capacity}, provide inaccurate benchmarks to the performance of short-packet communication systems, because of the assumption of asymptotically large blocklength~\cite{durisi16-09a,durisi16-02a}. 
In particular, these performance metrics are unable to capture the tension between the throughput gains in the transmission of short packets over wireless fading channels that are attainable by exploiting channel diversity, and the throughput losses caused by the insertion of pilot symbols, which are often used to estimate the wireless fading channel at the receiver~\cite{tong04-11a}.

A more useful performance metric for short-packet communication systems is the so called \emph{maximum coding rate} $\Rmax\lro{n,\epsilon}$, which is the largest rate achievable for a fixed blocklength~$n$, and a fixed packet error probability~$\epsilon$.
No closed-form expressions for $\Rmax\lro{n,\epsilon}$ are available for the channel models of interest in wireless communication systems. 
However, tight bounds on $\Rmax\lro{n,\epsilon}$ as well as second-order expansions in the limit $n\rightarrow\infty$ have been recently reported for a variety of wireless channel models.
These results rely on the nonasymptotic information-theoretic tools developed in~\cite{polyanskiy10-05a}. 

In this paper, we study the maximum coding rate achievable over Rician memoryless block-fading channels, for the case in which no \emph{a priori} \gls{csi} is available at the transmitter and at the receiver. 
Such a setup is of particular interest in sporadic short-packet transmissions subject to stringent latency constraints.
Indeed, the \gls{csi} that may have been acquired  at the receiver during previous packet transmissions is often outdated due to the sporadic nature of the transmissions, and  delay constraints may prevent the use of a feedback link, which is necessary for the transmitter to obtain \gls{csi}.
In practical wireless systems, the receiver typically obtains \gls{csi} through the use of \gls{pat} schemes~\cite{tong04-11a}, which involve multiplexing known pilot symbols among the data symbols within each packet. 
Our goal is to investigate the performance of such schemes when packets are short using a nonasymptotic information-theoretic analysis.

\subsection{Prior Art} 
\label{par:priorart}%
\paragraph*{The Nonfading AWGN Channel} 
\label{par:the_nonfading_awgn_channel}%
Tight upper (converse) and lower (achievability) bounds on $\Rmax(n,\epsilon)$ based on cone packing were obtained by Shannon~\cite{shannon59-a}.
Polyanskiy, Poor, and Verd\'u~\cite{polyanskiy10-05a} showed recently that Shannon's converse bound is a special case of the so-called \emph{min-max converse}~\cite[Thm.~27]{polyanskiy10-05a}, \cite{polyanskiy13-07b}, a general converse bound that involves a binary hypothesis test between the channel law and a suitably chosen auxiliary distribution. 
Furthermore, they obtained an alternative achievability bound---the $\kappa\beta$-bound~\cite[Thm.~25]{polyanskiy10-11a}---also based on binary hypothesis testing. 
This bound, although less tight than Shannon's achievability bound, is easier to evaluate numerically and to analyze asymptotically. 
Indeed, Shannon's achievability bound relies on the transmission of codewords that are uniformly distributed on the surface of an $\lro{n-1}$-dimensional  hypersphere in $\reals^n$ (a.k.a., \emph{spherical} or \emph{shell codes}), which makes the induced output distribution unwieldy. 
Min-max and $\kappa\beta$ bounds solve this problem by replacing the above-mentioned output distribution by a product Gaussian distribution, which is easier to analyze analytically.

Characterizing the min-max converse and the $\kappa\beta$ bound in the asymptotic regime of large blocklength~$n$, Polyanskiy, Poor, and Verd\'u established the following asymptotic expansion for $\Rmax(n,\epsilon)$ (see~\cite{polyanskiy10-05a} and also the refinement in~\cite{tan15-05a}), which, for convenience, we state  for the case of a \emph{complex} AWGN channel:
\begin{IEEEeqnarray}{rCL}\label{eq:normal_approximation}
  \Rmax(n,\epsilon)= C- \sqrt{n^{-1}V}Q^{-1}(\epsilon) +\landauO\lefto({n^{-1}\log n}\right).
\end{IEEEeqnarray}
Here, $C=\log(1+\rho)$, where $\rho$ denotes the SNR, is the channel capacity, $V=\rho(2+\rho)/(1+\rho)^2$ is the so-called channel \emph{dispersion}, $Q(\cdot)$ is the Gaussian $Q$ function, and $\landauO(n^{-1}\log n)$ comprises remainder terms of order $n^{-1}\log n$.

The expansion~\eqref{eq:normal_approximation}, which is commonly referred to as \emph{normal approximation} relies on a central-limit-theorem analysis and is accurate when $\Rmax$ is close to capacity.
When the target packet error probability is low and, hence, the maximum coding rate is far from capacity, large-deviation analyses resulting in the classic Gallager's \gls{rce}~\cite{gallager68a} yield more accurate results than~\eqref{eq:normal_approximation}.
%

\paragraph*{Fading Channels--no a-priori CSI} 
\label{par:fading_channels}
Bounds on $\Rmax$ for generic \emph{quasi-static multiple-antenna fading channels} were reported in~\cite{yang14-07c}.
Using these bounds, the authors showed that, under mild conditions on the probability distribution of the fading process, the channel dispersion (i.e., the parameter $V$ in~\eqref{eq:normal_approximation}) is zero. 
This means that the asymptotic limit (in this case the outage capacity) is approached much faster with~$n$ than in the AWGN case.
This is because the main source of error in quasi-static fading channels is the occurrence of ``deep fades'', which channel codes cannot mitigate.
The achievability bound in~\cite{yang14-07c} relies on a modified version of the $\kappa\beta$ bound, in which the decoder employs the following \emph{noncoherent} detection scheme: it computes the angle between the received signal and each one of the codewords, and picks the first codeword whose angle is smaller than a predetermined threshold.
The converse bound relies on the min-max converse~\cite[Thm.~27]{polyanskiy10-05a}.

The analysis in~\cite{yang14-07c} was later partly generalized in~\cite{durisi16-02a} to fading channels providing more than just a single diversity branch in time and/or frequency.
Specifically, the authors of~\cite{durisi16-02a} considered a multiantenna Rayleigh memoryless block-fading channel and assumed that coding can be performed across a fixed number of independently fading blocks.
The converse bound in~\cite{durisi16-02a} relies again on the min-max converse, whereas the achievability bound is built upon the so-called \gls{dt} bound~\cite[Thm.~17]{polyanskiy10-05a}. 
The input distribution used in~\cite{durisi16-02a} to compute the \gls{dt} bound is the one induced by \gls{ustm}~\cite{hochwald00-03a}, according to which the matrices describing the signal transmitted within each coherence block over the available transmit antennas are drawn independently from the uniform distribution on the set of unitary matrices and then they are scaled so as to satisfy the power constraint. 
This distribution, which achieves capacity at high SNR~\cite{zheng02-02a,yang13-02a} (provided that the sum of transmit and receive antennas does not exceed the length of the coherence block), corresponds---in the single-input single-output (SISO) case---to the transmission of independent shell codes over each coherence block.
Note that the resulting signaling scheme is noncoherent in that no pilot symbols are transmitted to learn the channel. 
Rather, information is conveyed through the choice of the subspace spanned by the row of each matrix, a quantity that is not affected by the fading. 
It is also worth remarking that the resulting bound assumes the adoption of an optimal receiver, able to compute the log-likelihood ratio of each codeword, which may be impractical.
The auxiliary distribution used in~\cite{durisi16-02a} to compute the min-max converse is the one induced by USTM.

Analyzing these achievability and converse bounds in the limit of both large SNR and large number of coherence blocks, the authors of\cite{lancho-serrano17-06a} obtained a simple-to-evaluate high-SNR normal approximation of the maximum coding rate $\Rmax$, which is in the same spirit as~\eqref{eq:normal_approximation}.
An attempt to analyze the scenario of imperfect \gls{csi} at the receiver for the case of \gls{mimo} Rayeigh block-fading channels was undertaken in~\cite{potter13}. The analysis, however, contains several inaccuracies.

For the multiple-antenna Rayleigh memoryless block-fading case, the input distribution achieving the \gls{rce} was studied by Abou-Faycal and Hochwald~\cite{abou-faycal99-a}.
They showed that it has the same structure as the ergodic-capacity-achieving input distribution~\cite{marzetta99-01a}, namely that the optimum input matrix is the product of a real, nonnegative diagonal matrix and an isotropically distributed unitary matrix. 
Furthermore, for the \gls{siso} case, they proved that for large SNR, the real-valued component becomes deterministic, and the input vector becomes a shell code.
The results in~\cite{abou-faycal99-a} were partly extended to single-antenna Rician memoryelss fading channels (coherence block of size one) in~\cite{gursoy06-a} where it is shown that the optimal scalar input has uniform phase and its amplitude is supported on  a finite number of mass-points.

An upper bound on the packet error probability based on the \gls{rce}  was derived in~\cite{ostman17-02a} for the \gls{mimo} case using \gls{ustm} as input distribution. 
Through numerical simulations, the authors showed that this bound is close to the one obtained in~\cite{durisi16-02a} using the \gls{dt} bound already at moderate error probabilities ($\epsilon\approx 10^{-4}$) in some scenarios.

\paragraph*{Pilot-Assisted Transmission and Mismatched Decoding} 
Analyses of \gls{pat} schemes in which the channel estimate is treated as perfect by a decoder that operates according to the \gls{nn} rule, fall into the general framework of mismatched decoding~\cite{kaplan93-a,merhav94-11a, ganti00-11a, lapidoth98-10a,lapidoth02-05a}.
A study of the performance of \gls{nn} decoders over fading channels under different assumptions on the availability of \gls{csi} was presented in~\cite{lapidoth02-05a}.
The analysis relies on using a Gaussian codebook and on the \gls{gmi}---an asymptotic quantity introduced in~\cite{kaplan93-a} that provides a lower bound on the maximum coding rate achievable for a fixed (possibly mismatched) decoding rule.\footnote{The authors of~\cite{kaplan93-a} analyze also the performance achievable over quasi-static Rician and Nakagami fading channels for the case of perfect \gls{csi} and no \gls{csi} with both matched and mismatched decoders, using the cut-off rate as asymptotic performance metric.}

Nonasymptotic lower bounds on the maximum coding rate achievable with mismatch decoding is presented in~\cite{scarlett14-05a} for the case of  \iid, constant-composition, and cost-constrained codes.
The analysis is based on the \gls{rcus}~\cite{martinez11-02a}, an adaptation and relaxation of the \gls{rcu} in~\cite{polyanskiy10-05a} for the case of mismatch decoder that recovers the generalized \gls{rce} introduced in~\cite{kaplan93-a}.


An analysis of the performance of \gls{pat} schemes using mutual information as asymptotic performance metric (and without imposing any restriction on the receiver structure) was carried out in~\cite{hassibi03-04a} for the case of \gls{mimo} Rayleigh block-fading channels.
%
It is shown that when one is allowed to optimize the power allocation between pilot and data symbols, it is optimal to use as many pilots per coherence block as the number of transmit antennas.
If instead pilot and data symbols need to be transmitted at the same power, the optimum number of pilots becomes SNR dependent, and a number of pilots much larger than the number of transmit antennas is needed in the low-SNR regime.
This investigation has been generalized to MIMO Rician-fading channels in~\cite{godavarti07-01a}.
Finally, a comprehensive asymptotic analysis of the performance of \gls{nn} decoders (and generalizations thereof) over MIMO fading channels using \gls{gmi} as performance metric can be found in~\cite{weingarten04-08a}.
\paragraph*{Channel codes for short packets} 
\label{par:codes_for_short_packets}
Recent surveys on the performance of actual coding schemes for short packet transmissions have been reported in, e.g.,~\cite{durisi16-09a,liva16a} for the case of AWGN channels.
The design of \gls{pat} schemes has been recently discussed in~\cite{liva17a} for the case of AWGN channel with deterministic unknown gain, and in~\cite{ostman17-02a} for the case of Rayleigh block-fading channels.
%

\subsection{Contributions} 
\label{par:contributions}
We study the maximum coding rate achievable over a \gls{siso} Rician memoryless block-fading channel under the assumption of no  \emph{a priori} \gls{csi}.
Specifically, we present converse and achievability bounds on the maximum coding rate that generalize and tighten the bounds previously reported in~\cite{durisi16-02a,Ostman17-03a}.
As in~\cite{durisi16-02a,Ostman17-03a} our converse bound relies on the min-max converse. 
Our two achievability bounds, which are built upon the \gls{rcus} bound, allow us to compare the performance of noncoherent and \gls{pat} schemes.
Specifically, the first bound relies on the transmission of \iid shell codes per coherence block and does not require explicit channel estimation at the receiver (while imposing no complexity constraint on the receiver architecture).
The second one, which has a more practical flavor and has not been analyzed before in the literature (including in our previous contribution~\cite{Ostman17-03a}), assumes \gls{pat} combined with shell codes for the transmission of the data symbols; furthermore, the receiver is constrained to perform \gls{ml} channel estimation based on the pilot symbols followed by \gls{nn} detection.

%

Through a numerical investigation, we show that our converse and achievability bounds delimit tightly the maximum coding rate, for a large range of SNR and Rician $\kappa$-factor values, and allow ones to identify---for given coding rate, packet size---the optimum number of coherence blocks to code over in order to minimize the energy per bit required to attain a target packet error probability.

Furthermore, our achievability bounds reveal that noncoherent transmission is more energy efficient than PAT even when the number of pilot symbols and their power is optimized. 
For example, for the case when a coded packet of $168$ symbols is transmitted using a channel code of rate $0.48$ bits/channel use over a Rayleigh block-fading channel with block size equal to $8$ symbols, the gap between the noncoherent and the PAT bound is about $1.2\dB$ at a packet error probability of~$10^{-3}$. 
This gap increases by a further $0.5\dB$ if pilot and data symbols are transmitted at the same power. 
When the power of the pilot symbols is optimized, one pilot symbol per coherence block turns out to suffice---a nonasymptotic counterpart of the result obtained in~\cite{hassibi03-04a}. 

We finally design an actual \gls{pat} scheme based on punctured tail-biting quasi-cyclic codes and a decoder that, using \emph{ordered statistics}, performs \gls{nn} detection based on \gls{ml} channel estimates.
The performance of this coding scheme is remarkably close to what predicted by our \gls{pat}-\gls{nn} achievability bound: $1\dB$ gap at $10^{-3}$ packet error probability for a packet of $168$ symbols, a code rate of $0.48$ bit/channel use, and transmission over a Rayleigh-fading channel with coherence block of $24$ symbols.  
This shows that our bound provides useful guidelines on the design of actual PAT schemes.
We also discuss how the performance of the decoder can be further improved (without hampering its relatively low computational complexity) by accounting for the inaccuracy of the channel estimates.

\paragraph*{Notation}
Uppercase letters such as $X$ and $\randvecx$ are used to denote scalar random variables and vectors, respectively; their realizations are written in lowercase, e.g., $x$ and $\vecx$.
The identity matrix of size $a\times a$ is written as $\matI_{a}$.
The distribution of a circularly-symmetric complex Gaussian random variable with variance $\sigma^2$ is denoted by $\cgauss{0}{\sigma^2}$. 
The superscript~$\tp{\lro{\cdot }}$ and $\herm{\lro{\cdot }}$ denote transposition and Hermitian transposition, respectively, and $\odot$ is the Schur product. 
Furthermore, $\veczero_{n}$ and $\vecone_{n}$ stand for the all-zero and all-one vectors of size $n$, respectively.
We write $\log\lro{\cdot}$ and $\log_2\lro{\cdot}$ to denote the natural logarithm and the logarithm to the base $2$, respectively.
Finally, $\lrh{a}^+$ stands for $\max\lrbo{0, a}$, we use $\Gamma\lro{\cdot}$ to denote the Gamma function, $I_{\nu}\lro{z}$ the modified Bessel function of the first kind, $\vecnorm{\cdot}$ the $l^2$-norm, and $\Ex{}{\cdot}$ the expectation operator.

\section{System Model} 
\label{sec:system_model}
We consider a \gls{siso} Rician memoryless block-fading channel.
Specifically, the random \gls{nlos} component is assumed to stay constant for $\nc$ successive channel uses (which form a coherence block) and to change independently across coherence blocks.
Coding is performed across $\ell$ such blocks; we shall refer to $\ell$ as the number of available \emph{diversity branches}.
The duration of each codeword (packet size) is, hence, $n=\nc\ell$.
This setup may be used to model, e.g., frequency-hopping systems and is relevant for  \gls{ofdm}-based systems (such as LTE and 5G), where a packet may consists of several resource blocks separated in frequency by more than the coherence bandwidth of the channel (see~\cite{ostman17-02a} for more details).
The \gls{los} component, i.e., the mean of the Rician fading random variable, which is assumed to be known at the receiver, stays constant over the duration of the entire packet (codeword).
No \emph{a priori} knowledge of the \gls{nlos} component is available at the receiver, in accordance to the no \emph{a priori} \gls{csi} assumption.

Mathematically, the channel input-output relation can be expressed as
\begin{IEEEeqnarray}{rCl}
\label{eq:sys_mod}
\randvecy_k =  H_k\vecx_k + \randvecw_k, \quad k = 1,\dots, \ell.
\end{IEEEeqnarray}
Here, $\vecx_k \in \complexset^{\nc}$ and $\randvecy_k \in \complexset^{\nc}$ contain the transmitted and received symbols within block~$k$, respectively. 
The Rician fading is modeled by $H_k \distas \cgauss{\mu\sub{H}}{\sigma^2\sub{H}}$ where $\mu\sub{H}=\sqrt{\kappa/(1+\kappa)}$ and  $\sigma^2\sub{H} = (1+\kappa)^{-1}$ with $\kappa$ being the Rician factor. 
Finally,  $\randvecw_k\distas \cgauss{\bm{0}}{\matI_{\nc}}$ is the AWGN noise. 
The random variables $\lrb{H_k}$ and $\lrb{\randvecw_k}$, which are mutually independent, are also independent over $k$. 

We next define a channel code.
\begin{dfn}
An $\lro{\ell, \nc, M, \epsilon, \rho}$-code for the channel (\ref{eq:sys_mod}) consists of
\begin{itemize}
\item 
An encoder $f:\lrbo{1, \dots, M} \rightarrow  \complexset^{\nc\ell}$ that maps the message $J$, which is uniformly distributed on $\lrbo{1,\dots, M}$ to a codeword in the set $\lrb{\vecc_1,\dots, \vecc_M}$.
Since each codeword $\vecc_m$, $m=1\dots, M$, spans $\ell$ blocks, it is convenient to express it as a concatenation of $\ell$ subcodewords of dimension~$\nc$
\begin{IEEEeqnarray}{rCl}
	\vecc_m = \lrho{\vecc_{m,1}, \dots, \vecc_{m,\ell}}.
\end{IEEEeqnarray}
We require that each subcodeword satisfies the average-power constraint
\begin{IEEEeqnarray}{rCl}\label{eq:power_constraint}
\vecnorm{\vecc_{m,k}}^2 =  \nc\rho, \quad k=1,\dots, \ell.
\end{IEEEeqnarray}
Since the noise has unit variance, we can think of $\rho$ as the average SNR per symbol.

\item A decoder $g: \complexset^{\nc\ell} \rightarrow \lrbo{1,\dots, M}$ satisfying an average error probability constraint
\begin{IEEEeqnarray}{rCl}\label{eq:avPrC}
\frac{1}{M} \sum_{j=1}^M \Pr\lrbo{g\lro{\randvecy^\ell} \neq J \given J=j}\leq \epsilon
\end{IEEEeqnarray}
where $\randvecy^\ell = \lrho{\randvecy_1, \dots, \randvecy_\ell}$ is the channel output induced by the codeword $\vecx^\ell = \lrho{\vecx_1, \dots, \vecx_\ell} = f(j)$.
\end{itemize}
\end{dfn}
For given $\ell$ and $n_c$, $\epsilon$, and $\snr$, the maximum coding rate \Rmax, measured in information bits per channel use, is defined as
\begin{IEEEeqnarray}{rCl}
\label{eq:maximal_rate}
R^*(\ell,\nc,\epsilon,\rho) = \sup\lrbo{\frac{\log_2 M}{\ell\nc} : \exists \lro{\ell, \nc, M, \epsilon, \rho}\!\text{-code}}.
\end{IEEEeqnarray}
In words, for a fixed blocklength $\ell\nc$ and a fixed SNR $\rho$, we seek the largest number $M^*$ of codewords that can be transmitted with average error probability not exceeding $\epsilon$.
The maximum coding rate is then given by $R^*=(\log_2 M^*)/(\ell\nc)$.

In practical applications, we are often interested in the problem of minimizing the SNR $\rho$ for a fixed packet error probability, a fixed blocklength $\ell\nc$, and a fixed number of information bits $\log_2 M$.
This yields the following alternative optimization problem:
\begin{equation}\label{eq:min_eps}
  \rho^*(\ell,\nc,M,\epsilon)= \inf\lrbo{\rho : \exists \lro{\ell, \nc, M, \epsilon, \rho}\!\text{-code}}.
\end{equation}
Throughout, we will repeatedly use that upper and lower bounds on $R^*$ can be translated into lower and upper bounds on $\snr^*$ and vice versa.
Also, we will often express our results in terms of the minimum energy per bit $\Emin$, which is related to $\rho^*$ as 
\begin{equation}\label{eq:minimum_energy_per_bit}
  \frac{E\sub{b}^*}{N_0}(\ell,\nc,M,\epsilon) = \frac{\ell\nc}{\log_2 M}\,\rho^*(\ell,\nc,M,\epsilon).
\end{equation}
%

\section{Finite-blocklength bounds on \Rmax} 
We shall next present achievability and converse bounds on $\Rmax$ obtained by using the nonasymptotic information-theoretic tools developed in~\cite{polyanskiy10-05a,martinez11-02a}.
In Section~\ref{sec:a_noncoherent_lower_bound_on_rmax} we provide an achievability bound that is based on the \gls{rcus}~\cite[Thm.~1]{martinez11-02a} and on the use of \iid shell codes, as input distribution, across the coherence blocks.
This bound does not require an explicit estimation of the fading channel at the receiver.
Rather, it relies on a \emph{noncoherent} transmission technique in which the message is encoded in the direction of the input vectors $\{\vecx_k\}$ in~\eqref{eq:sys_mod}--a quantity that is not affected by the fading process.

In Section~\ref{sec:pilots_NN}, we provide a second achievability bound, which relies instead on \gls{pat}.
We assume that the receiver uses pilot symbols to obtain a \gls{ml} estimate of the channel fading (we do not assume the fading law to be known at the receiver), which is then fed to a \gls{nn} decoder that treats it as perfect.
This bound relies once more on the \gls{rcus}; furthermore, \iid shell codes across the coherence blocks are used in the channel uses dedicated to the data symbols.

Since both bounds cannot be expressed in closed form and require Monte-Carlo simulation for their numerical evaluations (which may be time consuming for low $\epsilon$ values), we present also easy-to-evaluate relaxations of these two bounds based on the generalized \gls{rce}.

In order to investigate the potential gains attainable by using a \gls{pat} scheme in which the receiver is aware of the channel distribution, and accounts for the imperfect nature of the \gls{csi}, we develop in Section~\ref{sec:pilots_ML} a \gls{pat}-based achievability bound, where knowledge of the joint distribution between the fading process and its (pilot-based) estimate allows the decoder to operate according to the \gls{ml} principle.
This bound tightens the one presented in~\cite{Ostman17-03a}.

Finally, in Section~\ref{sec:a_general_upper_bound_on_rmax}, we present a converse bound on $\Rmax$ that relies on the min-max converse~\cite[Thm.~27]{polyanskiy10-05a}, with auxiliary distribution chosen as the distribution of $\lrbo{\randvecy_k}$ induced by the transmission of independent shell codes over each coherence block.
This bound generalizes to Rician-fading channels the one presented in~\cite{durisi16-02a} for the Rayleigh-fading case.

\subsection{Achievability Bounds on \Rmax: Preliminaries} 
\label{sec:lower_bound_on_rmax}
Throughout the paper, we shall assume that the decoder produces an estimate $\widehat{m}$ of the transmitted message as follows:  
\begin{equation}\label{eq:decoder}
  \widehat{m}= \argmax_{m} q^\ell\lro{\vecc_m,\vecy^\ell}.
\end{equation}
Here, $\{\vecc_m\}_{m=1}^M$ are the codewords and $\vecy^\ell$ is the received signal.
Furthermore,
\begin{equation}\label{eq:decoding_metric}
  q^{\ell}(\vecx^\ell,\vecy^\ell)=\prod_{k=1}^{\ell}q(\vecx_k,\vecy_k)
\end{equation}
where $q(\vecx_k,\vecy_k)$ is a bounded nonnegative function, which we refer to as \emph{decoding metric}.
In the next sections we will introduce the decoding metrics that are relevant for our achievability results. 
Before doing so, we review the \gls{rcus} bound and its connections to the generalized \gls{rce}.
\begin{thm}[\gls{rcus} bound~{\cite[Th.~1]{martinez11-02a}}]\label{thm:rcus}
For every input distribution $P_{\randvecx^{\ell}}$ and every decoding metric $q\lro{\cdot, \cdot}$, there exists a $\lro{\ell, \nc, M, \epsilon, \rho}$-code with decoder operating according to~\eqref{eq:decoder} and with average-error probability upper-bounded as
\begin{IEEEeqnarray}{rCl}\label{eq:rcus-original}
\epsilon \leq \text{\gls{rcus}}\lro{\ell,\nc,M,\rho} =\inf_{s\geq 0} \Ex{}{e^{-\lrho{i^{\ell}_s\lro{\randvecx^{\ell}, \randvecy^{\ell}} - \log\lro{M-1}}^+}} 
\end{IEEEeqnarray}
where 
\begin{equation}\label{eq:generalized_inf_dens}
    i^{\ell}_s\lro{\vecx^{\ell}, \vecy^{\ell}} = \log \frac{q^{\ell}\lro{\vecx^{\ell}, \vecy^{\ell}}^s}{\Ex{}{q^{\ell}\lro{\randvecx^{\ell}, \vecy^{\ell}}^s}}
\end{equation}
is the \emph{generalized information density}.
\end{thm}
%

Assume now that the input distribution factorizes as
\begin{equation}\label{eq:block-memoryless-input}
  P_{\randvecx^{\ell}}(\vecx^{\ell})=\prod_{k=1}^\ell P_{\randvecx}(\vecx_k)
\end{equation}
i.e., the vector $\randvecx^\ell=[\randvecx_1,\dots \randvecx_{\ell}]$ has \iid $\nc$-dimensional components $\{\randvecx_k\}$ all distributed according to $P_{\randvecx}$.
It follows from~\eqref{eq:decoding_metric} that the generalized information density in~\eqref{eq:generalized_inf_dens} can be rewritten as 
\begin{IEEEeqnarray}{rCL}\label{eq:gen_inf_dens_block_additive}
  i^{\ell}_s\lro{\vecx^{\ell}, \vecy^{\ell}}=\sum_{k=1}^{\ell}\log\frac{q\lro{\vecx_{k}, \vecy_{k}}^s}{\Ex{}{q\lro{\randvecx_k, \vecy_k}^s}}=\sum_{k=1}^{\ell} i_s(\vecx_k,\vecy_k).
\end{IEEEeqnarray}
Let now
\begin{equation}\label{eq:gallager function for mismatch decoding}
  E_0\lro{\tau, s} = -\log\Ex{}{e^{-\tau\,  i_s\lro{\randvecx, \randvecy}}}
\end{equation}
be the Gallager's function for mismatch decoding~\cite{kaplan93-a}.
Here, $(\randvecx, \randvecy)\distas P_{\randvecx}P_{\randvecy\given \randvecx}$, where $P_{\randvecy\given \randvecx}$ is the channel law (within a coherence block) corresponding to the input-output relation~\eqref{eq:sys_mod}. 
Furthermore, fix a rate $R>0$ (measured for convenience in nats per channel use) and let
\begin{equation}\label{eq:generalized_random_coding_error_exponent}
  E\lro{\nc,R,\rho}=\sup_{s\geq 0, \tau\in\lrho{0,1}} \lrbo{E_0\lro{\tau, s} - \tau \nc R}
\end{equation}
be the generalized \gls{rce}. 
It follows from~\cite{martinez11-02a} that 
\begin{equation}\label{eq:generalized_error_exp_from_rcus}
  E\lro{\nc,R,\rho}=\sup_{s\geq 0} \lim_{\ell \rightarrow \infty} -\frac{1}{\ell} \log \lro{\text{\gls{rcus}}\lro{\ell,\nc,2^{\ell\nc R},\rho}}.
\end{equation}
In words, for fixed $\nc,R,\rho$, the \gls{rcus} bound decays to zero exponentially fast in $\ell$, with exponent given by the generalized \gls{rce}.
An application of a Chernoff-type bound yields the following classic achievability bound based on the generalized \gls{rce}. 
This bound is less tight than the \gls{rcus} bound in Theorem~\ref{thm:rcus} but it is often easier to evaluate numerically.
%
\begin{cor}[generalized \gls{rce} bound]\label{col:grce} 
For every $P_{\randvecx}$ in~\eqref{eq:block-memoryless-input} and every decoding metric $q(\cdot,\cdot)$ there exists a $\lro{\ell, \nc, M, \epsilon, \rho}$-code with decoder operating according to~\eqref{eq:decoder} and with average-error probability upper-bounded as
\begin{IEEEeqnarray}{rCl} 
\epsilon \leq e^{-\ell E\lro{\nc,R,\rho}}
\end{IEEEeqnarray}
where $R=(\log M)/(\nc\ell)$.
\end{cor}

\subsection{Noncoherent Achievability Bound on \Rmax} 
\label{sec:a_noncoherent_lower_bound_on_rmax}
To derive our noncoherent achievability bound, we set
\begin{equation}\label{eq:ML_decoder}
  q(\vecx_k,\vecy_k)=P_{\randvecy\given\randvecx}(\vecy_k\given\vecx_k).
\end{equation}
It follows then from~\eqref{eq:decoding_metric} and~\eqref{eq:decoder} that the corresponding decoder operates according to the \gls{ml} rule.
Furthermore, we take $P_{\randvecx}$ in~\eqref{eq:block-memoryless-input} to be a shell distribution, i.e., the uniform distribution over all vectors $\vecx \in \complexset^{\nc}$ satisfying the power constraint  $\vecnorm{\vecx}^2=\nc\snr$ (cf.~\eqref{eq:power_constraint}).
With these choices, the \gls{rcus} bound in Theorem~\ref{thm:rcus}, applied to the channel~\eqref{eq:sys_mod}, takes the following form.
%
\begin{thm}[\gls{rcus} noncoherent achievability bound]
\label{thm:RCUs_ML}
The maximum coding rate~\Rmax in~\eqref{eq:maximal_rate} achievable over the channel~\eqref{eq:sys_mod} is lower-bounded as
\bie
	R^*(\ell,\nc,\epsilon,\rho) \geq \max\lrbo{\frac{\log_2\lro{M}}{\nc\ell} : \epsilon\sub{ub}\lro{\ell,\nc,M,\rho} \leq \epsilon}
\eie
where 
\begin{equation}
 \label{eq:rcus_ML}
	 \epsilon\sub{ub}\lro{\ell,\nc,M,\rho} = \inf_{s\geq 0}\Ex{}{\exp\lefto\{-\lrho{ \sum_{k=1}^\ell S^s_k - \log\lro{M-1} }^+\right\}}
\end{equation}
with 
\bie
\label{eq:Sk}
	S_k^s & = & \lro{\nc-2}\log\lro{s}  - \log\lro{\frac{1+\sigma\sub{H}^2\nc\rho  }{\sigma\sub{H}^{2}}} - \log\lro{\Gamma\lro{\nc}} \nonumber \\ 
	& &  -  s\lro{\vecnorm{\randvecw_k}^2-\vecnorm{\widetilde{\randvecw}_k}^2} + \frac{s\abs{\mu\sub{H}}^2}{\sigma\sub{H}^2}  -\log \int_{\positivereals} \frac{  \exp\lro{-s\lro{\rho\nc + \sigma\sub{H}^{-2}} z }}{  \lro{\vecnorm{\widetilde{\randvecw}_k} \sqrt{\rho \nc z}}^{\nc-1} } \nonumber \\
	& & \times 
  I_{\nc-1}\lro{2s\vecnorm{\widetilde{\randvecw}_k}\sqrt{ \rho \nc z}}
  	  I_{0}\lro{2s\sigma\sub{H}^{-2}\sqrt{{z\abs{\mu\sub{H}}^2}}}  \mathrm{d}z.
\eie
Here, the $\{\randvecw_k\}$ are defined as in \eqref{eq:sys_mod} and
\bie
\label{eq:wide_tilde}
\widetilde{\randvecw}_k =\begin{bmatrix} \mu\sub{H} \sqrt{\nc\rho} \\ \veczero_{\nc-1} \end{bmatrix} +\begin{bmatrix} \sqrt{\sigma\sub{H}^2 \nc \rho + 1} \\ \vecone_{\nc-1} \end{bmatrix} \odot \randvecw_k.
\eie
\end{thm}

\begin{IEEEproof}
See Appendix \ref{app:RCUs_proof}.
\end{IEEEproof}

By setting $\mu\sub{H}=0$, $\sigma\sub{H}^2=1$, and $s=1$ in~\eqref{eq:Sk} and~\eqref{eq:wide_tilde}, one recovers a \gls{siso} version of the achievability bound reported in~\cite[Th.~1]{durisi16-02a} for the Rayleigh-fading case.
The bound in~\cite[Th.~1]{durisi16-02a} does not involve an optimization over the parameter $s$ because it is based on the \gls{dt} bound, which is less tight than the \gls{rcus} bound and coincides with it when $s=1$.

Note that the expectation in~\eqref{eq:rcus_ML} is not known in closed form, which makes the numerical evaluation of the bound demanding, especially for low values of $\epsilon$.
We next present an alternative noncoherent lower bound on $R^*$ obtained by relaxing the \gls{rcus} to the  \gls{rce} in  Corollary~\ref{col:grce}. 
Although less tight than the bound in Theorem~\ref{thm:RCUs_ML}, the resulting bound is easier to evaluate numerically.


\begin{cor}[\gls{rce} noncoherent achievability bound]\label{thm:RCE_ML}
The maximum coding rate~\Rmax in~\eqref{eq:maximal_rate} achievable over the channel~\eqref{eq:sys_mod} is lower-bounded as
\bie
	\Rmax\lro{\ell,\nc,\epsilon,\rho} \geq \max \lrbo{\frac{\log_2\lro{M}}{\nc\ell} : \epsilon\sub{ub}\lro{\ell,\nc,M,\rho} \leq \epsilon}
\eie
where 
\begin{equation}
	 \epsilon\sub{ub}\lro{\ell,\nc,M,\rho} = e^{-\ell E\lro{\nc,R,\rho}}
\end{equation}
with  
$R=(\log M)/(\nc\ell)$ and
\begin{IEEEeqnarray}{rCl} 
 E\lro{\nc,R,\rho}  =  \max_{0 \leq \tau \leq 1} \lrbo{E_0\lro{\tau} - \tau\nc R}.
\end{IEEEeqnarray}
Here, 
\bie
	E_0\lro{\tau} &&= -\log\lro{c\lro{\tau} \int_{0}^{\infty} r^{\nc-1}e^{-r} J(r,\tau)^{1+\tau}\mathrm{d}r}\label{eq:gallager_function_noncoherent}
\eie
where
\bie
c\lro{\tau} &=&  \lro{1+\sigma\sub{H}^2\rho\nc}^\tau \Gamma\lro{\nc}^{\tau}  e^{-\abs{\mu_{H}}^2 /\sigma_{H}^{2} } \lrho{\frac{\lro{1+\tau}^{\nc-2}}{ \sigma_{H}^2}} ^{1+\tau}\label{eq:constant_c}
\eie
and
\bie
J(r,\tau) &=& \int_{0}^{\infty} \frac{   e^{-\frac{1}{1+\tau}\lr{ \sigma_{H}^{-2} +  \rho\nc} z}   }{ \lr{ \sqrt{r\rho\nc z }}^{\nc-1}  }  I_{\nc-1} \lro{\frac{2\sqrt{r \rho \nc  z}}{1+\tau}}I_0\lro{\frac{2\abs{\mu_{H}}\sqrt{ z}}{\sigma_{H}^{2}\lro{1+\tau}}} \mathrm{d}z.\label{eq:function_c}
\eie
\end{cor}

\begin{IEEEproof}
See Appendix \ref{app:RCE_ML_proof}.
\end{IEEEproof}

By setting $\mu\sub{H}=0$ and $\sigma\sub{H}^2=1$ in~\eqref{eq:constant_c} and~\eqref{eq:function_c}, one recovers a \gls{siso} version of the \gls{rce} bound reported in~\cite[Th.~3]{ostman17-02a} for the Rayleigh-fading case.


\subsection{Pilot-Assisted Nearest-Neighbor Achievability Bound on $\Rmax$} \label{sec:pilots_NN}
We  assume that, within each coherence block, $\np$ out of the available $\nc$ channel uses are reserved for pilot symbols.
The remaining $\nd=\nc-\np$ channel uses convey the data symbols.
We  further assume that all pilot symbols are transmitted at power $\rp$, and that the data symbol vectors $\vecx_k^{(\text{d})}\in \complexset^{\nd}$  satisfy the power constraint $\vecnorm{\vecx_k^{(\text{d})}}^2=\nd \rd$, $k=1,\dots,\ell$. 
We require that $\np\rp + \nd\rd = \nc \rho$ so as to fulfill~\eqref{eq:power_constraint}.


The receiver uses the $\np$ pilot symbols available in each coherence block to perform a \gls{ml} estimation of the corresponding fading coefficient. 
Specifically, for a given pilot vector $\vecx_k^{(\text{p})}$ and a corresponding received-signal vector $\vecy_k^{(\text{p})}$, the receiver computes the estimate 
\begin{equation}\label{eq:ml_fading_estimation}
 \widehat{h}_k= \herm{\bigl(\vecx_k^{(\text{p})}\bigr)}\vecy_k^{(\text{p})}/{\vecnorm{\vecx_k^{(\text{p})}}}^{2}.
\end{equation}
It follows from~\eqref{eq:ml_fading_estimation} that, given $H_k=h_k$, we have 
$\widehat{H}_k \sim \cgauss{h_k}{1/(\np \rp)}$.

We further assume that the fading estimate $\hat{h}_k$ is fed to a \gls{nn} detector that treats it as perfect. 
Specifically, we consider the following decoding metric:
\begin{IEEEeqnarray}{rCl} \label{eq:nn_metric}
   q\lro{\vecx_k, \vecy_k} = e^{-\vecnorm{\vecy_k^{(\text{d})} - \widehat{h}_k \vecx_k^{(\text{d})}}^2}
\end{IEEEeqnarray}
where $\hat{h}_k$ is computed as in~\eqref{eq:ml_fading_estimation}.
Finally, we take as input distribution $P_{\randvecx^{\text{d}}}$  the uniform distribution over all vectors $\vecx \in \complexset^{\nd}$ satisfying  $\vecnorm{\vecx}^2=\nd\rd$.

Under these assumptions, the \gls{rcus} bound in Theorem~\ref{thm:rcus} takes the following form.

\begin{thm}[\gls{rcus}--\gls{pat}--\gls{nn} achievability bound]
\label{thm:RCUs_NN}
Fix two nonnegative integers $\np$ $(\np<\nc)$ and $\nd=\nc-\np$, and two nonnegative real-valued parameters  $\rp$ and $\rd$ satisfying $\np\rp + \nd\rd = \nc \rho$.
The maximum coding rate~\Rmax in~\eqref{eq:maximal_rate} achievable over the channel~\eqref{eq:sys_mod} is lower-bounded as
\bie
	\Rmax\lro{\ell,\nc,\epsilon,\rho} \geq \max\lrbo{\frac{\log_2\lro{M}}{\nc\ell} : {\epsilon\sub{ub}}\lro{\ell,\nc,M,\rho}\leq \epsilon}
\eie
where 
\begin{equation}\label{eq:rcus-pat-snn-epsilon-bound}
	 {\epsilon}\sub{ub}\lro{\ell,\nc,M,\rho} = 
   \min_{s\geq 0}\Ex{}{\exp\lefto\{-\lrho{ \sum_{k=1}^\ell T^s_k - \log\lro{M-1} }^+\right\}}
\end{equation}
where
\bie
\label{eq:Tk}
	T_k^s & = & s\lro{\vecnorm{\overline{\randvecw}_k}^2 - \vecnorm{\widetilde{\randvecw}_k}^2  }+ s\nd\rd\abs{\widehat{H}_k}^2 - \log{\Gamma\lro{\nd}} \nonumber\\
	&& + \lro{\nd-1}\log\lro{s\abs{\widehat{H}_k}\vecnorm{\overline{\randvecw}_k}\sqrt{\nd\rd} }   - \log\lro{I_{\nd-1}\lro{2s\abs{\widehat{H}_k}\vecnorm{\overline{\randvecw}_k}\sqrt{\nd\rd} }}.
\eie
Here, 
\bie
\label{eq:wide_tilde}
\overline{\randvecw}_k =\begin{bmatrix} H_k \sqrt{\nd\rd} \\ \veczero_{\nd-1} \end{bmatrix} +  \randvecw_k 
\quad \text{and} \quad
\widetilde{\randvecw}_k = \begin{bmatrix} \sqrt{ \nd \rd/(\np\rp) + 1} \\ \vecone_{\nd-1} \end{bmatrix} \odot \randvecw_k 
\eie
with $\randvecw_k \sim \cgauss{\veczero_{\nd}}{\matI_{\nd}}$.
The expectation in~\eqref{eq:rcus-pat-snn-epsilon-bound} is with respect to the joint distribution $\prod_{k=1}^{\ell} P_{ H_k, \widehat{H}_k, \randvecw_k}$ 
where 
$ P_{ H_k, \widehat{H}_k, \randvecw_k}=
P_{H_k} P_{\widehat{H}_k \given H_k} P_{\randvecw_k} $ 
with 
$P_{H_k}=\jpg(\mu\sub{H},\sigma^2\sub{H})$ and $P_{\widehat{H}_k\given H_k=h}=\jpg(h,1/(\np\rp))$.

\end{thm}

\begin{IEEEproof}
See Appendix \ref{app:RCUs_NN_proof}.
\end{IEEEproof}

As in Section \ref{sec:a_noncoherent_lower_bound_on_rmax}, we present an alternative, easier-to-compute achievability bound, which is obtained by relaxing the \gls{rcus} used in Theorem~\ref{thm:RCUs_NN} to the generalized \gls{rce} in Corollary~\ref{col:grce}.
\begin{cor}[\gls{rce}--\gls{pat}--\gls{nn} achievability bound]\label{thm:RCE_NN}
Fix two nonnegative integers $\np$ $(\np<\nc)$ and $\nd=\nc-\np$, and two nonnegative real-valued parameters  $\rp$ and $\rd$ satisfying $\np\rp + \nd\rd = \nc \rho$.
The maximum coding rate~\Rmax in~\eqref{eq:maximal_rate} achievable over the channel~\eqref{eq:sys_mod} is lower-bounded as
\bie
	\Rmax\lro{\ell,\nc,\epsilon,\rho} \geq \max \lrbo{\frac{\log_2\lro{M}}{\nc\ell} : \epsilon\sub{ub}\lro{\ell,\nc,M,\rho} \leq \epsilon}
\eie
where 
\begin{equation}\label{eq:error_exponent_pat-nn}
	 \epsilon\sub{ub} \lro{\ell,\nc,M,\rho} =  \Ex{}{  e^{-\ell E(\nc,R,\rho, \widehat{H})}}
\end{equation}
with  
$R=(\log M)/(\nc\ell)$ and
where the expectation is with respect to~$P_{\widehat{H}} = \cgauss{\mu\sub{H}}{\sigma\sub{H}^2+1/(\np\rp)}$.
The error exponent $E(\nc,R,\rho, \widehat{h})$ is 
\begin{IEEEeqnarray}{rCl} 
E(\nc,R,\rho, \widehat{h})  =   \max_{0\leq\tau\leq 1} \max_{s > 0}\lrbo{E_0(\tau, s, \widehat{h}) - \tau \nc R}
\end{IEEEeqnarray}
and the Gallager's function for mismatch decoding $E_0(\tau, s, \widehat{h})$ is 
\begin{IEEEeqnarray}{rCl}
 E_0(\tau, s, \widehat{h})  &=& -\log c(\widehat{h})  \int_{0}^{\infty} r^{\nd-1} e^{-r}   J(r,\tau, s, \widehat{h}) \mathrm{d}r 
\end{IEEEeqnarray}
where $c(\widehat{h}) =   \sigma\sub{p}^{-2} \mathrm{exp}\lro{ - \frac{\abs{\mu\sub{p}(\widehat{h})}^2 \rd\nd}{ 1 + \sigma\sub{p}^2 \rd \nd}}$
with
\begin{equation}
 \mu\sub{p}(\widehat{h}) = \frac{\sigma\sub{H}^2 \widehat{h} + (\np\rp)^{-1} \mu\sub{H}}{\sigma\sub{H}^2+(\np\rp)^{-1}}, \quad \sigma\sub{p}^2 = \frac{\sigma\sub{H}^2(\np\rp)^{-1}}{\sigma\sub{H}^2+(\np\rp)^{-1}}\label{eq:parameters_eq_channel}.
\end{equation}
Furthermore,
\begin{IEEEeqnarray}{rCl} 
J\bigl(r,\tau, s, \widehat{h}\bigr) &=&  \frac{\Gamma\lro{\nd} ^\tau I_{\nd-1}(2s\abs{\widehat{h}}\sqrt{r \rd \nd })^{\tau}} {(s\abs{\widehat{h}}\sqrt{r \rd \nd })^{\tau\lro{\nd - 1}}} \exp\lro{\abs{a(\widehat{h})}^2 \lro{\frac{\rd\nd}{1+\sigma\sub{p}^{2}\rd\nd} - \frac{1}{\sigma\sub{p}^{2}} }}\nonumber\\
 && \times  \int_{0}^{\infty} \frac{\exp\lro{-\lro{ \sigma\sub{p}^{-2} + \rd\nd}z}}{ \lro{\sqrt{r z \rd \nd}}^{\nd-1} }  
  I_{\nd-1}\lro{2\sqrt{r z \rd \nd}} I_0(2\abs{a(\widehat{h})} \sigma\sub{p}^{-2}  \sqrt{z} ) \mathrm{d}z \IEEEeqnarraynumspace
\end{IEEEeqnarray} 
with $a(\widehat{h}) = \mu\sub{p}(\widehat{h}) - \widehat{h}s\tau \lro{ 1 + \sigma\sub{p}^2 \rd \nd}$.
\end{cor}
\begin{IEEEproof}
See Appendix \ref{app:RCE_NN_proof}.
\end{IEEEproof}

\subsection{Pilot-Assisted Maximum Likelihood Achievability Bound on \Rmax} \label{sec:pilots_ML}
To assess the performance loss due to the (mismatch) \gls{nn} decoding metric~\eqref{eq:nn_metric}, we present next a \gls{pat}-based achievability bound in which this metric is replaced by the ML metric
\begin{equation}\label{eq:ml-pat-decoding-metric}
  q(\vecx_k,\vecy_k)=P_{\randvecy^{(\text{d})}\given \randvecx^{(\text{d})},\hat{H}}(\vecy^{\text{(d)}}_k\given \vecx^{\text{(d)}}_k,\hat{h}_k)
\end{equation}
where $\hat{h}_k$ is the ML channel estimate~\eqref{eq:ml_fading_estimation}.
As argued in the proof of Corollary~\ref{thm:RCE_NN}, 
\begin{equation}\label{eq:conditional_channel_given_estimate}
  P_{\randvecy^{(\text{d})}\given \randvecx^{(\text{d})},\hat{H}}(\vecy^{\text{(d)}}_k\given \vecx^{\text{(d)}}_k,\hat{h}_k)=\cgauss{ \mu\sub{p}(\widehat{h}_k) \vecx^{\text{(d)}}_k}{\sigma\sub{p}^2 \vecx^{\text{(d)}}_k\herm{(\vecx^{\text{(d)}}_k)}+\matI_{\nd}}
\end{equation}
where $\mu\sub{p}(\widehat{h}_k)$ and $\sigma\sub{p}^2 $ are defined in~\eqref{eq:parameters_eq_channel}.
This implies that, given the channel estimate $\hat{h}_k$ and the input vector $\vecx^{\text{(d)}}_k$, the conditional \gls{pdf} of $\randvecy^{\text{(d)}}_k$ coincides with the law of the following channel
\begin{equation}\label{eq:io_relation_pilots}
  	\randvecy_k^{\lro{\text{d}}} = Z_k\vecx_k^{\lro{\text{d}}}  + \randvecw_k, \quad k=1,\dots,\ell.
\end{equation}
Here, $Z_k\distas\cgauss{\mu\sub{p}(\widehat{h}_k)}{\sigma\sub{p}^2}$ and $\randvecw_k\distas \cgauss{\veczero_{\nd}}{\matI_{\nd}}$.

%
%
%
We see from~\eqref{eq:io_relation_pilots} that we can account for the availability of the noisy \gls{csi} $\{\widehat{H}_k=\widehat{h}_k\}$ simply by transforming the Rician fading channel~\eqref{eq:sys_mod} into the equivalent Rician fading channel~\eqref{eq:io_relation_pilots}, whose LOS component is a random variable that depends on the channel estimates $\{\widehat{H}_k\}$.
A lower bound on $\Rmax$ for this setup can be readily obtained by assuming that each $\nd$-dimensional data vector is generated independently from a shell code, by applying Theorem~\ref{thm:RCUs_ML} to each realization of $\{\widehat{H}_k\}$, and then by averaging over $\{\widehat{H}_k\}$. 
%
%
%
%
\begin{thm}[\gls{rcus}--\gls{pat}--\gls{ml} achievability bound]\label{thm:RCUs_Pilot}
Fix two nonnegative integers $\np$ $(\np<\nc)$ and $\nd=\nc-\np$, and two nonnegative real-valued parameters  $\rp$ and $\rd$ satisfying $\np\rp + \nd\rd = \nc \rho$.
The maximum coding rate~\Rmax in~\eqref{eq:maximal_rate} achievable over the channel~\eqref{eq:sys_mod} is lower-bounded as
\begin{IEEEeqnarray}{rCl}
	\Rmax\lro{\ell,\nc,\epsilon,\rho} \geq \max\lrbo{\frac{\log_2\lro{M}}{\nc\ell} : \epsilon\sub{ub}\lro{\ell,\nc,M,\rho} \leq \epsilon}
\end{IEEEeqnarray}
where 
\begin{IEEEeqnarray}{rCl}\label{eq:rcus-pat-ml-epsilon-bound}
	 \epsilon\sub{ub}\lro{\ell,\nc,M,\rho} =  \min_{s\geq 0} \Ex{}{\exp\lefto\{-\lrho{ \sum_{k=1}^\ell \bar{S}^s_k(\widehat{H}_k) - \log\lro{M-1} }^+\right\}}.
\end{IEEEeqnarray} 
The expectation in~\eqref{eq:rcus-pat-ml-epsilon-bound} is with respect to  $\prod_{k=1}^{\ell} P_{ \widehat{H}_k} P_{\randvecw_k}$ where $P_{ \widehat{H}_k}  = \cgauss{\mu\sub{H}}{\sigma\sub{H}^2 + (\np\rp)^{-1} }$ and $P_{\randvecw_k}\distas\cgauss{\veczero_{\nd}}{\matI_{\nd}}$. 
The random variables $\{\bar{S}^s_k(\widehat{H}_K)\}$ are defined similarly as in~\eqref{eq:Sk} 
with the difference that $\nc$, $\rho$, $\mu\sub{H}$ and $\sigma\sub{H}^2$ in~\eqref{eq:Sk} are 
replaced by $\nd$, $\rd$, $\mu\sub{p}(\widehat{H_k})$ and $\sigma\sub{p}^2$, respectively.
\end{thm}
\vspace*{2mm}

For the case $\np=0$, the pilot-based achievability bound in Theorem~\ref{thm:RCUs_Pilot} coincides 
with the noncoherent bound given in Theorem~\ref{thm:RCUs_ML}.
Furthermore, by setting $\rho\sub{d} = 
\rho\sub{p}$ and $s=1$, we recover \cite[Th.~3]{Ostman17-03a}.\footnote{With $(M-1)/2$ replaced by $M-1$.}
The bound in Theorem~\ref{thm:RCUs_Pilot} can be relaxed to a generalized-\gls{rce}-type bound by proceeding as in the proof of Corollary~\ref{thm:RCE_ML}.

\subsection{A Converse Bound on \Rmax} 
\label{sec:a_general_upper_bound_on_rmax}


We next state our converse bound.\footnote{This bound was first presented in the conference version of this paper~\cite[Th.~2]{Ostman17-03a}.}
\begin{thm}[Min-max converse bound]
\label{thm:MC}
The maximum coding rate~\Rmax in~\eqref{eq:maximal_rate} achievable over the channel~\eqref{eq:sys_mod} is upper-bounded as
\begin{equation}
\label{eq:MC}
 R^*\leq \inf_{\lambda \geq 0} \frac{1}{\ell\nc} \lro{\lambda - \log\lrho{\Pr\lrbo{\sum_{k=1}^\ell S_k^1 \leq \lambda} - \epsilon}^+}
\end{equation}
where the random variables $ \lrbo{S_k^1}_{k=1}^\ell$ are obtained by setting $s=1$ in~\eqref{eq:Sk}.
\end{thm}
\begin{IEEEproof}
See Appendix \ref{app:MC_proof}.
\end{IEEEproof}

By setting $\mu_{H}=0$ and $\sigma^2_{H}=1$, one recovers a \gls{siso} version of the min-max converse bound obtained in~\cite{durisi16-02a} for the Rayleigh-fading case.

\section{Numerical Results} 
\label{sec:numerical_results}

\subsection{Dependency of \Rmax and $E\sub{b}^*/N_0$ on the Rician Factor $\kappa$ } 
\label{sec:numerical_results_a}

\begin{figure}
\centering
 \begin{tikzpicture}
\pgfplotsset{
    scaled y ticks = false,
    width=\fwidth*0.45,
    height=\fwidth*0.4,
     title style={yshift=-6pt,}
}
    \begin{groupplot}[ 
        group style={
        group size=3 by 4,
        vertical sep=45pt,
        horizontal sep=35pt,
       },
    ]
    \nextgroupplot[   axis x line*=bottom,
         xmode = log,
         xlabel={Number of diversity branches $\ell$ (log scale)},
         ylabel = Bit/channel use,
         xmin = 2,
         xmax = 84,
         ymin = 0,
         xticklabels={{$2$}, {$4$}, {$7$}, {$14$}, {$21$},{$28$}, {$42$},{$84$}},
        xtick={2, 4 , 7,14,21,28,42, 84},
        xlabel near ticks,
        grid=both   ]
         \addplot[color = black, ,mark size=1.5,line width=0.5mm, forget plot] table [y index={1}, col sep=comma] {./Data/Gaussian_ref.csv};
         
      \addplot[name path = p1,color = blue, solid,mark=*,mark size=\marksize] table [y index={1}, col sep=comma] {./Data/RICE_USTM_INDUCED1x1_6dB_kappa_1000_n_168.csv}coordinate[pos=0.07](ut1);
         \addplot[name path = p2, mark=square*, color = red, solid,mark size=\marksize] table [y index={2}, col sep=comma] {./Data/RICE_USTM_INDUCED1x1_6dB_kappa_1000_n_168.csv}coordinate[pos=0.07](pt1);
         \addplot[name path = p3,color = orange, solid,mark=triangle*,mark size=\marksize] table [y index={5}, col sep=comma] {./Data/E_r_Rate_vs_kappa_n_168_1e-3.csv};
	 \addplot[gray,fill opacity=0.2] fill between[of=p2 and p1];
    
     \coordinate (pt2) at ($(pt1) !.5! (ut1)$);
     \draw (pt2) ellipse  (2pt and 7pt);
     \coordinate (pt3) at ($(pt2)+ (0pt,-7pt)$);
     \coordinate (pt4) at ($(pt3)+ (+5pt,-14pt)$);
     \draw[<-] (pt3)--(pt4) node at ($(pt4) + (13pt,-2pt)$) {$\kappa=10^3$};
     
%
%
       \addplot[name path = p1,color = blue, solid,mark=*,mark size=\marksize,forget plot] table [y index={1}, col sep=comma] {./Data/RICE_USTM_INDUCED1x1_6dB_kappa_10_n_168.csv}coordinate[pos=0.78](ut1);
         \addplot[name path = p2,color = red, solid,mark=square*,mark size=\marksize,forget plot] table [y index={2}, col sep=comma] {./Data/RICE_USTM_INDUCED1x1_6dB_kappa_10_n_168.csv}coordinate[pos=0.77](pt1);
         \addplot[name path = p3,color = orange, solid,mark=triangle*,mark size=\marksize,forget plot] table [y index={3}, col sep=comma] {./Data/E_r_Rate_vs_kappa_n_168_1e-3.csv};
 \addplot[gray,fill opacity=0.2] fill between[of=p2 and p1];
     \coordinate (pt2) at ($(pt1) !.5! (ut1)$);
     \draw (pt2) ellipse  (2pt and 7pt);
     \coordinate (pt3) at ($(pt2)+ (0pt,-7pt)$);
     \coordinate (pt4) at ($(pt3)+ (+5pt,-14pt)$);
     \draw[<-] (pt3)--(pt4) node at ($(pt4) + (11pt,-2pt)$) {$\kappa=10$};
     
%

     \addplot[name path = p1, color = blue, solid,mark=*,mark size=\marksize,forget plot] table [y index={1}, col sep=comma] {./Data/RICE_USTM_INDUCED1x1_6dB_kappa_0_n_168.csv}coordinate[pos=0.85](ut1);
         \addplot[name path = p2, color = red, solid,mark=square*,mark size=\marksize,forget plot] table [y index={1}, col sep=comma] {./Data/RCUs_ML_RvNc_SNR_6_n_168_np_0_kappa_0.csv}coordinate[pos=0.2](pt1);
         \addplot[color = orange, solid,mark=triangle*,mark size=\marksize,forget plot] table [y index={1}, col sep=comma] {./Data/E_r_Rate_vs_kappa_n_168_1e-3.csv};
 \addplot[gray,fill opacity=0.2] fill between[of=p2 and p1];
     \draw ($(ut1) + (0pt,-3pt)$)  ellipse  (2pt and 6pt);
     \coordinate (pt3) at ($(ut1) + (0pt,3pt)$);
     \coordinate (pt4) at ($(pt3) + (5pt,8pt)$);
     \draw[<-] (pt3)--(pt4) node at ($(pt4) + (8pt,4pt)$) {$\kappa=0$};
\coordinate (c2) at (rel axis cs:1,1);

    \nextgroupplot[at=(group c1r1.west),
              xmin = 2,
       xmax = 84,
      xlabel = Size of coherence block $n_c$,
       xmode = log,
       xticklabels={{$84$}, {$42$}, {$24$}, {$12$}, {$8$}, {$6$},{$4$},{$2$}},
       xtick={2,  4, 7, 14,  21, 28, 42,84},
       hide y axis,
       axis x line*=top,
       xlabel near ticks]
\addplot[draw = none] table [y index={1}, col sep=comma] {./Data/Gaussian_ref.csv};
    \nextgroupplot[    
     	 axis x line*=bottom,
         xmode = log,
         xlabel={Number of diversity branches $\ell$ (log scale)},
         ylabel = $ E\sub{b}/N_0 \> \lrho{\text{dB}} $,
         xmin = 2,
         xmax = 84,
         ymin = 0,
         ymax = 15,
         xticklabels={{$2$}, {$4$}, {$7$}, {$14$}, {$21$},{$28$}, {$42$},{$84$}},
        xtick={2, 4 , 7,14,21,28,42, 84},
        xlabel near ticks,
        grid=both,
        legend style={at={($(0,0)+(1cm,1cm)$)},legend columns=4,fill=none,draw=black,anchor=center,align=center},
        legend to name=fred
    ]
	      	      
       \addplot[color = black, ,mark size=1.5,line width=0.5mm] table [y index={1}, col sep=comma] {./Data/Normal_approx_EbN0.csv}coordinate[pos=0.3](pnorm);\addlegendentry{Normal approximation}
        
      \addplot[name path = p1, color = blue, solid,mark=*,mark size=\marksize] table [y index={1}, col sep=comma] {./Data/EbN0_vs_Kappa_Converse.csv}coordinate[pos=0.78](ut1);\addlegendentry{Min-max converse	}
      \addplot[name path = p2, color = red, solid,mark=square*,mark size=\marksize] table [y index={1}, col sep=comma] {./Data/EbN0_vs_Kappa_RCUs.csv};\addlegendentry{\gls{rcus} noncoherent}
     \addplot[ color = orange, solid,mark=triangle*,mark size=\marksize] table [y index={1}, col sep=comma] {./Data/RCE_EbN0_vs_kappa_n_168_1e-3.csv};\addlegendentry{\gls{rce} noncoherent}
     
      \addplot[gray,fill opacity=0.2] fill between[of=p2 and p1];
     \draw ($(ut1) + (0pt,5pt)$)  ellipse  (2pt and 9pt);
     \coordinate (pt3) at ($(ut1) + (0pt,14pt)$);
     \coordinate (pt4) at ($(pt3) + (5pt,8pt)$);
     \draw[<-] (pt3)--(pt4) node at ($(pt4) + (8pt,4pt)$) {$\kappa=0$};
     \addplot[name path = p1, color = blue, solid,mark=*,mark size=\marksize,forget plot] table [y index={3}, col sep=comma] {./Data/EbN0_vs_Kappa_Converse.csv}coordinate[pos=0.15](ut1);
     \addplot[name path = p2, color = red, solid,mark=square*,mark size=\marksize,forget plot] table [y index={3}, col sep=comma] {./Data/EbN0_vs_Kappa_RCUs.csv};
     \addplot[color = orange, solid,mark=triangle*,mark size=\marksize,forget plot] table [y index={3}, col sep=comma] {./Data/RCE_EbN0_vs_kappa_n_168_1e-3.csv};
      
     \addplot[gray,fill opacity=0.2] fill between[of=p2 and p1];
     \draw ($(ut1) + (0pt,8pt)$)  ellipse  (2pt and 15pt);
     \coordinate (pt3) at ($(ut1) + (0pt,23pt)$);
     \coordinate (pt4) at ($(pt3) + (5pt,8pt)$);
     \draw[<-] (pt3)--(pt4) node at ($(pt4) + (8pt,4pt)$) {$\kappa=10$};
	 \addplot[name path = p1, color = blue, solid,mark=*,mark size=\marksize,forget plot] table [y index={5}, col sep=comma] {./Data/EbN0_vs_Kappa_Converse.csv}coordinate[pos=0.02](ut1);
	 \addplot[name path = p2, color = red, solid,mark=square*,mark size=\marksize,forget plot] table [y index={5}, col sep=comma] {./Data/EbN0_vs_Kappa_RCUs.csv};
     \addplot[color = orange, solid,mark=triangle*,mark size=1.5,forget plot] table [y index={5}, col sep=comma] {./Data/RCE_EbN0_vs_kappa_n_168_1e-3.csv};
     
      \addplot[gray,fill opacity=0.2] fill between[of=p2 and p1];
     \draw ($(ut1) + (0pt,3pt)$)  ellipse  (2pt and 6pt);
     \coordinate (pt3) at ($(ut1) + (0pt,9pt)$);
     \coordinate (pt4) at ($(pt3) + (5pt,4pt)$);
     \draw[<-] (pt3)--(pt4) node at ($(pt4) + (8pt,4pt)$) {$\kappa=10^3$};

    \coordinate (c1) at (rel axis cs:0,1);
    
    \nextgroupplot[at=(group c3r1.north),
       xmin = 2,
       xmax = 84,
       xlabel = Size of coherence block $n_c$,
       xmode = log,
       xticklabels={{$84$}, {$42$}, {$24$}, {$12$}, {$8$}, {$6$},{$4$},{$2$}},
       xtick={2,  4, 7, 14,  21, 28, 42,84},
       hide y axis,
       axis x line*=top,
       xlabel near ticks]
       
\addplot [draw=none] table [y index={1}, x index = {0}, col sep=comma] {./Data/R_vs_nc_MC_DT_n_168_SNR_6_eps_1e-3.csv};

\end{groupplot}
\coordinate (c3) at ($(c1)!.5!(c2)$);
\node[above] at (c3 |- current bounding box.north)
{\pgfplotslegendfromname{fred}};
\node[text width=6cm,align=center,anchor=north,font=\footnotesize] at ([yshift=-5mm]group c1r2.south) {\subcaption{$E\sub{b}^*/N_0$ for $R=0.48$ bit/channel use. \label{fig:1a}}};
\node[text width=6cm,align=center,anchor=north,font=\footnotesize] at ([yshift=-5mm]group c1r1.south) {\subcaption{\Rmax for $\rho=6$ dB. \label{fig:1b}}};
\end{tikzpicture}%
     \caption{\gls{rcus} noncoherent achievability bound (Theorem~\ref{thm:RCUs_ML}), its \gls{rce} relaxation  (Corollary~\ref{thm:RCE_ML}), and min-max converse (Theorem~\ref{thm:MC});  $\kappa=\lrb{0,10,1000}$, $\epsilon = 10^{-3}$ and $n = 168$.}
 \label{fig:1}
 \end{figure}
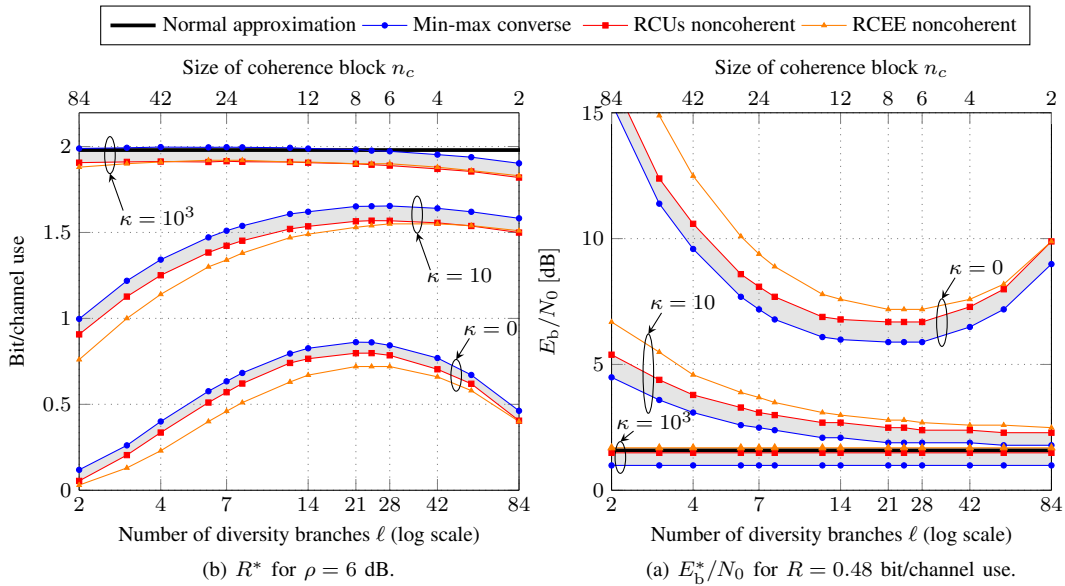  

 In Fig.~\ref{fig:1}, we plot the \gls{rcus} noncoherent achievability bound
 (Theorem~\ref{thm:RCUs_ML}), its \gls{rce} relaxation (Corollary~\ref{thm:RCE_ML}), and the min-max converse bound (Theorem~\ref{thm:MC}).
 We assume a blocklength of $n=168$ channel uses and a packet error probability of $\epsilon=10^{-3}$.
In Fig.~\ref{fig:1a}, we set $\rho=6\dB$ and investigate the dependency of $R^*$ on the number of 
diversity branches $\ell$ or, equivalently, on the size of each coherence block $\nc$.
In Fig.~\ref{fig:1b}, we investigate instead, for a fixed rate $R=0.48$ bit/channel use (and,
 hence, a fixed number of information bits, since $n=168$), the minimum energy 
per bit ${E\sub{b}^*}/{N_0}$ in~\eqref{eq:minimum_energy_per_bit} needed to achieve $\epsilon=10^{-3}$.
 
 We see from Fig.~\ref{fig:1} that the bounds are tight and allow one to identify
the optimal number of diversity branches that maximizes \Rmax or, equivalently, minimizes ${E\sub{b}^*}/{N_0}$. 
 For $\kappa=0$ (Rayleigh-fading) this number is $\ell^*\approx 21$.
 When $\ell<\ell^*$, the performance bottleneck is the limited diversity available. 
 When $\ell>\ell^*$, the limiting factor is instead the fast channel variations (which manifest themselves in a small coherence block $\nc$).
 We note also that, as $\kappa$ increases, both \Rmax and ${E\sub{b}^*}/{N_0}$ become less sensitive to $\ell$.
 This is expected since, when $\kappa\to \infty$, the Rician channel converges to a nonfading AWGN channel.
 Indeed, we see that the bounds obtained for the case $\kappa=10^{3}$ are in good agreement with the normal approximation~\eqref{eq:normal_approximation}.
Note also that the agreement with the normal approximation is better for smaller values of $\ell$.
This is because, in the AWGN case, the optimum input distribution involves shell codes over $\complexset^{n}$, whereas our bounds rely on shell codes over $\complexset^{\nc}$. 

 %
 As expected, the \gls{rcus} bound is tighter that the \gls{rce} bound, which is however easier to evaluate numerically. 
\subsection{PAT or Noncoherent?}
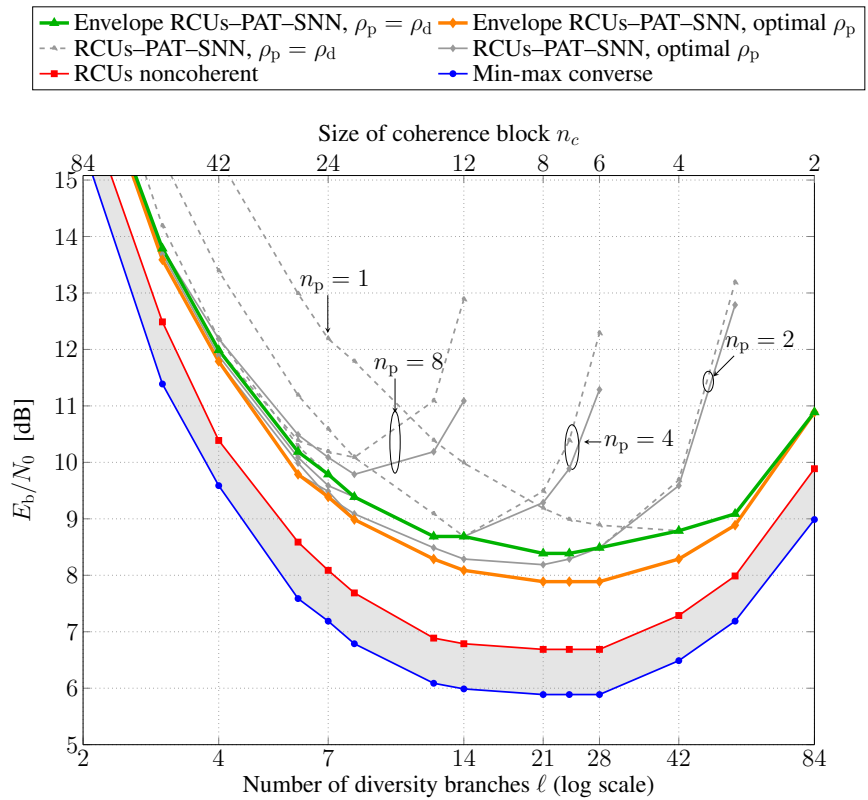
\begin{figure}[t]
  \centering
  \resizebox{0.7\textwidth}{!}{
  \begin{tikzpicture}[font=\large]
 \begin{axis}
     [
       axis x line*=bottom,
       xmode = log,
          xlabel={Number of diversity branches $\ell$ (log scale)},
         ylabel = {$E\sub{b}/N_0 \>$  [dB]},
         xmin = 2,
         xmax = 84,
         ymin = 5,
         ymax = 15.08,
         xticklabels={{$2$}, {$4$}, {$7$}, {$14$}, {$21$},{$28$}, {$42$},{$84$}},
         ylabel style={ yshift=3ex},
        xtick={2, 4 , 7,14,21,28,42, 84},
      xlabel near ticks,
      grid=both,
      legend style={at={(0.5,1.15)},anchor=south},
      reverse legend,
      legend columns=2,
      legend cell align=left,
     ]

 \addplot [name path = MC, color = blue, mark=*,solid, mark size=1.5,line width=\linew] table [y index={1}, x index = {0}, col sep=comma] {./Data/RCUs_ML_SNRvsL_Converse_R_0.48_n_168_np_0_kappa_0_epsilon_1e-03.csv};\addlegendentry{Min-max converse}

 \addplot [name path = DT, color = red, mark=square*,solid, mark size=1.5,line width=\linew] table [y index={1},x index = {0},  col sep=comma] {./Data/RCUs_ML_SNRvsL_R_0.48_n_168_np_0_kappa_0_epsilon_1e-03.csv};\addlegendentry{\gls{rcus} noncoherent}
 \addplot [color=gray, opacity=0.2, forget plot] fill between[of=MC and DT];

 \addplot[mark=diamond*,solid, mark size=1.5,line width=\linew, color = gray!80] table [y index={1},x index = {0},  col sep=comma] {./Data/RCUs_NN_SNRvsL_R_0.48_n_168_np_1_kappa_0_epsilon_1e-03boost.csv};\addlegendentry{\gls{rcus}--\gls{pat}--\gls{nn}, optimal $\rp$ }
 \addplot[ mark=diamond*,solid, mark size=1.5,line width=\linew, color = gray!80, forget plot] table [y index={1},x index = {0},  col sep=comma] {./Data/RCUs_NN_SNRvsL_R_0.48_n_168_np_2_kappa_0_epsilon_1e-03boost.csv};
 \addplot[ mark=diamond*,solid, mark size=1.5,line width=\linew, color = gray!80, forget plot] table [y index={1},x index = {0},  col sep=comma] {./Data/RCUs_NN_SNRvsL_R_0.48_n_168_np_4_kappa_0_epsilon_1e-03boost.csv}coordinate[pos=0.1](pt1);
 \addplot[ mark=diamond*,solid, mark size=1.5,line width=\linew, color =  gray!80, forget plot] table [y index={1},x index = {0},  col sep=comma] {./Data/RCUs_NN_SNRvsL_R_0.48_n_168_np_8_kappa_0_epsilon_1e-03boost.csv};

 \addplot[mark=triangle*,dashed, mark size=1.5 ,line width=\linew, color = gray!80] table [y index={1},x index = {0},  col sep=comma] {./Data/RCUs_NN_SNRvsL_R_0.48_n_168_np_1_kappa_0_epsilon_1e-03.csv}coordinate[pos=0.6](ut1);\addlegendentry{\gls{rcus}--\gls{pat}--\gls{nn}, $\rp=\rd$}

 \addplot[ mark=triangle*,dashed, mark size=1.5,line width=\linew, color = gray!80, forget plot] table [y index={1},x index = {0},  col sep=comma] {./Data/RCUs_NN_SNRvsL_R_0.48_n_168_np_2_kappa_0_epsilon_1e-03.csv}coordinate[pos=0.9](ut2);

 \addplot[ mark=triangle*,dashed, mark size=1.5,line width=\linew, color = gray!80, forget plot] table [y index={1},x index = {0},  col sep=comma] {./Data/RCUs_NN_SNRvsL_R_0.48_n_168_np_4_kappa_0_epsilon_1e-03.csv}coordinate[pos=0.86](ut4);


 \addplot[ mark=triangle*,dashed, mark size=1.5,line width=\linew, color = gray!80, forget plot] table [y index={1},x index = {0},  col sep=comma] {./Data/RCUs_NN_SNRvsL_R_0.48_n_168_np_8_kappa_0_epsilon_1e-03.csv}coordinate[pos=0.78](ut8);

 \addplot[ mark=diamond*,solid,mark options={solid,fill=orange}, mark size=1.5,line width=2pt, color = orange] table [y index={1},x index = {0},  col sep=comma] {./Data/Envelope_NN_boost.csv};\addlegendentry{Envelope \gls{rcus}--\gls{pat}--\gls{nn}, optimal $\rp$ }
 
 \addplot[ mark=triangle*,solid, mark size=1.5,line width=2pt, color = black!30!green] table [y index={1},x index = {0},  col sep=comma] {./Data/Envelope_NN.csv};\addlegendentry{Envelope \gls{rcus}--\gls{pat}--\gls{nn}, $\rp=\rd$}


     \coordinate (pt3) at ($(ut1) + (0pt,3pt)$);
     \coordinate (pt4) at ($(ut1) + (0pt,25pt)$);
     \draw[<-] (pt3)--(pt4) node at ($(pt4) + (4pt,8pt)$) {$\np=1$};
     
          \draw ($(ut2) + (0pt,-3pt)$)  ellipse  (3pt and 6pt);
     \coordinate (pt3) at ($(ut2) + (3pt,0pt)$);
     \coordinate (pt4) at ($(ut2) + (15pt,12pt)$);
     \draw[<-] (pt3)--(pt4) node at ($(pt4) + (15pt,8pt)$) {$\np=2$};
     
          \draw ($(ut4) + (2pt,-3pt)$)  ellipse  (4pt and 13pt);
     \coordinate (pt3) at ($(ut4) + (9pt,0pt)$);
     \coordinate (pt4) at ($(pt3) + (10pt,0pt)$);
     \draw[<-] (pt3)--(pt4) node at ($(pt4) + (22pt,0pt)$) {$\np=4$};
     
          \draw ($(ut8) + (0pt,-8pt)$)  ellipse  (3pt and 18pt);
     \coordinate (pt3) at ($(ut8) + (0pt,9pt)$);
     \coordinate (pt4) at ($(pt3) + (0pt,20pt)$);
     \draw[<-] (pt3)--(pt4) node at ($(pt4) + (8pt,8pt)$) {$\np=8$};

\end{axis}

  \begin{axis}
 [
       xmin = 2,
       xmax = 84,
      xlabel = Size of coherence block $n_c$,
       xmode = log,
       xticklabels={{$84$}, {$42$}, {$24$}, {$12$}, {$8$}, {$6$},{$4$},{$2$}},
       xtick={2,  4, 7, 14,  21, 28, 42,84},
       hide y axis,
       axis x line*=top,
       xlabel near ticks
 ]
     \addplot[draw=none] table [y index={1}, x index = {0}, col sep=comma] {./Data/R_vs_nc_MC_DT_n_168_SNR_6_eps_1e-3.csv};

     \end{axis}
  \end{tikzpicture}
  }
  \caption{ $E\sub{b}^*/N_0$ for $n=168$, $\epsilon=10^{-3}$ and $R=0.48$ bit/channel use; 
   min-max converse (Theorem \ref{thm:MC}), \gls{rcus} noncoherent achievability bound (Theorem 
   \ref{thm:RCUs_ML}), and \gls{rcus}--\gls{pat}--\gls{nn} achievability bound (Theorem 
   \ref{thm:RCUs_NN}). 
   The dashed lines are obtained by assuming $\rd=\rp$;  the solid lines are obtained by optimizing over the power allocation.}
 \label{fig:EbN0vsL_NN}
 \end{figure}

 In Fig. \ref{fig:EbN0vsL_NN}, we compare the \gls{rcus} noncoherent achievability bound (Theorem~\ref{thm:RCUs_ML}) with the \gls{rcus}--\gls{pat}--\gls{nn} achievability bound (Theorem~\ref{thm:RCUs_NN}). 
This last bound is computed for different numbers of pilot symbols $\np$.
We consider both the case in which pilot and data symbols are transmitted at the same power ($\rp=\rd$) and the case in which the power allocation is optimized. 
The min-max converse (Theorem~\ref{thm:MC}) is also depicted for reference.
The parameters are the same as in Fig.~\ref{fig:1}: $n=168$, $\epsilon=10^{-3}$, $R=0.48$ bit/channel use.
Furthermore, we assume $\kappa=0$.
For the case $\rp=\rd$, we see that the optimum number of pilot symbols decreases as the size $\nc$ of the coherence block decreases, as expected. 
Indeed, when the coherence block is small, the rate penalty resulting for increasing the number of pilot symbols overcomes the rate gain resulting from the more accurate channel estimation. 
When one performs an optimization over the power allocation, however, one pilot symbol per coherence block suffices (the curve for $\np=1$ overlaps with the corresponding envelope in Fig. \ref{fig:EbN0vsL_NN}).
This is in agreement with what proven in~\cite[Th.~3]{hassibi03-04a} using mutual information as asymptotic performance metric. 
Furthermore, the optimum power allocation turns out to follow closely the asymptotic rule provided in~\cite[Th.~3]{hassibi03-04a}.

We see from Fig.~\ref{fig:EbN0vsL_NN} that, when $\ell=28$, the gap between the \gls{rcus} noncoherent bound and the \gls{rcus}--\gls{pat}--\gls{nn}  bound with optimum power allocation is about $1.2\dB$.
This gap increases further by $0.6\dB$ if the additional constraint $\rp=\rd$ is imposed.

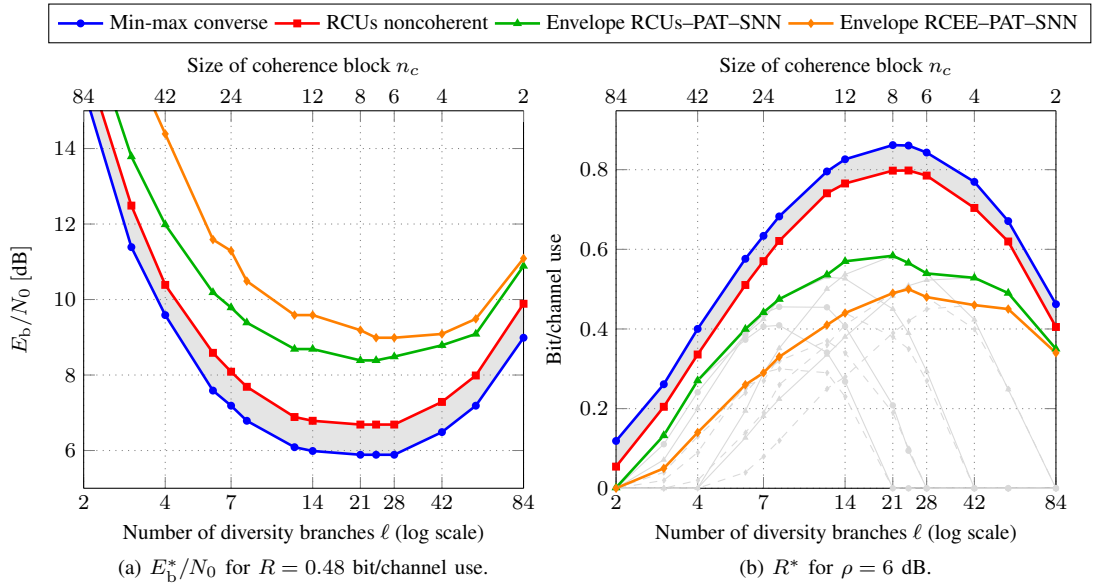
\begin{figure}[h!]
\centering
 \begin{tikzpicture}
\pgfplotsset{
    scaled y ticks = false,
    width=\fwidth*0.45,
    height=\fwidth*0.4,
     title style={yshift=-6pt,}
}
    \begin{groupplot}[ 
        group style={
        group size=3 by 4,
        vertical sep=45pt,
        horizontal sep=35pt
        },
    ]
    \nextgroupplot[    
     axis x line*=bottom,
       xmode = log,
          xlabel={Number of diversity branches $\ell$ (log scale)},
         ylabel = $E\sub{b}/N_0 \> \lrho{\text{dB}}$,
         xmin = 2,
         xmax = 84,
         ymin = 5,
         ymax = 15,
         xticklabels={{$2$}, {$4$}, {$7$}, {$14$}, {$21$},{$28$}, {$42$},{$84$}},
        xtick={2, 4 , 7,14,21,28,42, 84},
        xlabel near ticks,
        grid=both,
        legend style={at={($(0,0)+(1cm,1cm)$)},legend columns=4,fill=none,draw=black,anchor=center,align=center},
        legend to name=fred
    ]
 \addplot [name path = MC, color = blue, mark=*,solid, mark size=\marksize,line width=\linew] table [y index={1}, x index = {0}, col sep=comma] {./Data/RCUs_ML_SNRvsL_Converse_R_0.48_n_168_np_0_kappa_0_epsilon_1e-03.csv};\addlegendentry{Min-max converse}
 \addplot [name path = DT, color = red, mark=square*,solid, mark size=\marksize,line width=\linew] table [y index={1},x index = {0},  col sep=comma] {./Data/RCUs_ML_SNRvsL_R_0.48_n_168_np_0_kappa_0_epsilon_1e-03.csv};\addlegendentry{\gls{rcus} noncoherent}
 \addplot [color=gray, opacity=0.2, forget plot] fill between[of=MC and DT];
 \addplot[ mark=triangle*,solid, mark size=\marksize,line width=\linew, color = black!30!green] table [y index={1},x index = {0},  col sep=comma] {./Data/Envelope_NN.csv};\addlegendentry{Envelope \gls{rcus}--\gls{pat}--\gls{nn}}
  \addplot[ mark=diamond*,solid, mark size=\marksize,line width=\linew, color = orange] table [y index={5},x index = {0},  col sep=comma] {./Data/RCE_NN_Envelope_EbN0_vs_L.csv};\addlegendentry{Envelope \gls{rce}--\gls{pat}--\gls{nn}}

    \coordinate (c1) at (rel axis cs:0,1);
    
    \nextgroupplot[at=(group c1r1.west),
              xmin = 2,
       xmax = 84,
      xlabel = Size of coherence block $n_c$,
       xmode = log,
       xticklabels={{$84$}, {$42$}, {$24$}, {$12$}, {$8$}, {$6$},{$4$},{$2$}},
       xtick={2,  4, 7, 14,  21, 28, 42,84},
       hide y axis,
       axis x line*=top,
       xlabel near ticks]
\addplot [draw=none] table [y index={1}, x index = {0}, col sep=comma] {./Data/R_vs_nc_MC_DT_n_168_SNR_6_eps_1e-3.csv};
    \nextgroupplot[   axis x line*=bottom,
       xmode = log,
          xlabel={Number of diversity branches $\ell$ (log scale)},
         ylabel = Bit/channel use,
         xmin = 2,
         xmax = 84,
         ymin = 0,
         xticklabels={{$2$}, {$4$}, {$7$}, {$14$}, {$21$},{$28$}, {$42$},{$84$}},
        xtick={2, 4 , 7,14,21,28,42, 84},
		legend pos=north west,
      xlabel near ticks,
      grid=both   ]
 \addplot [name path = MC, color = blue, solid,mark=*,mark options={solid, mark size=\marksize},line width=\linew] table [y index={1}, x index = {0}, col sep=comma] {./Data/R_vs_nc_MC_DT_n_168_SNR_6_eps_1e-3.csv}coordinate[pos=0.4](p1);
 \addplot [name path = DT, color = red, solid,mark=square*,mark options={solid, mark size=\marksize},line width=\linew] table [y index={1},x index = {0},  col sep=comma] {./Data/RCUs_ML_RvNc_SNR_6_n_168_np_0_kappa_0.csv}coordinate[pos=0.38](p1);
 \addplot [color=gray, opacity=0.2, forget plot] fill between[of=MC and DT];
 \addplot[dashed,mark=diamond*,color = gray!30, mark size=\marksize] table [y index={2},x index = {0},  col sep=comma] {./Data/R_vs_nc_Er_NN_PAT_n_168_SNR_6_eps_1e-3.csv}coordinate[pos=0.25](p1);
 %
 \addplot [dashed,mark=diamond*,color = gray!30, mark size=\marksize] table [y index={3},x index = {0},  col sep=comma] {./Data/R_vs_nc_Er_NN_PAT_n_168_SNR_6_eps_1e-3.csv};
 %
 \addplot [ dashed,mark=diamond*, color = gray!30, mark size=\marksize] table [y index={4},x index = {0},  col sep=comma] {./Data/R_vs_nc_Er_NN_PAT_n_168_SNR_6_eps_1e-3.csv};
 %
 \addplot [dashed,mark=diamond*, color = gray!30, mark size=\marksize] table [y index={5},x index = {0},  col sep=comma] {./Data/R_vs_nc_Er_NN_PAT_n_168_SNR_6_eps_1e-3.csv};
 %
 \addplot [dashed,mark=diamond*,color = gray!30, mark size=\marksize] table [y index={6},x index = {0},  col sep=comma] {./Data/R_vs_nc_Er_NN_PAT_n_168_SNR_6_eps_1e-3.csv}coordinate[pos=0.52](p2);

       \addplot[color = gray!30,mark = triangle*, mark size=\marksize] table [x index ={0},y index={6}, col sep=comma] {./Data/Envelope_NN_R_vs_L.csv};   
        \addplot[color = gray!30,mark = triangle*, mark size=\marksize ] table [x index ={0},y index={7}, col sep=comma] {./Data/Envelope_NN_R_vs_L.csv};    
        \addplot[color = gray!30,mark = triangle*, mark size=\marksize] table [x index ={0},y index={9}, col sep=comma] {./Data/Envelope_NN_R_vs_L.csv};    
        \addplot[ color = gray!30,mark = *, mark size=\marksize] table [x index ={0},y index={11}, col sep=comma] {./Data/Envelope_NN_R_vs_L.csv};   
        \addplot[ color = gray!30,mark = *, mark size=\marksize ] table [x index ={0},y index={13}, col sep=comma] {./Data/Envelope_NN_R_vs_L.csv};    
 
\addplot[color = green,mark = triangle*, line width = \linew, color = black!30!green, mark size=\marksize] table [x index ={0},y index={5}, col sep=comma] {./Data/Envelope_NN_R_vs_L.csv};

\addplot[mark = diamond*, line width = \linew, color = orange, mark size=\marksize] table [x index ={0},y index={6}, col sep=comma] {./Data/RCE_NN_Envelope.csv};

\coordinate (c2) at (rel axis cs:1,1);

    \nextgroupplot[at=(group c3r1.north),
              xmin = 2,
       xmax = 84,
      xlabel = Size of coherence block $n_c$,
       xmode = log,
       xticklabels={{$84$}, {$42$}, {$24$}, {$12$}, {$8$}, {$6$},{$4$},{$2$}},
       xtick={2,  4, 7, 14,  21, 28, 42,84},
       hide y axis,
       axis x line*=top,
       xlabel near ticks]
  \addplot [draw=none] table [y index={1}, x index = {0}, col sep=comma] {./Data/R_vs_nc_MC_DT_n_168_SNR_6_eps_1e-3.csv};
\end{groupplot}
\coordinate (c3) at ($(c1)!.5!(c2)$);
\node[above] at (c3 |- current bounding box.north)
{\pgfplotslegendfromname{fred}};

\node[text width=6cm,align=center,anchor=north,font=\footnotesize] at ([yshift=-5mm]group c1r1.south) {\subcaption{$E\sub{b}^*/N_0$ for $R=0.48$ bit/channel use. \label{fig:NN_comp_a}}};
\node[text width=6cm,align=center,anchor=north,font=\footnotesize] at ([yshift=-5mm]group c1r2.south) {\subcaption{\Rmax for $\rho=6$ dB. \label{fig:NN_comp_b}}};
\end{tikzpicture}%
  \caption{Comparison between \gls{rcus}-\gls{pat}--\gls{nn} (Theorem \ref{thm:RCUs_NN}) and \gls{rce}-\gls{pat}--\gls{nn} (Corollary \ref{thm:RCE_NN}) for $\kappa=0$, $n=168$, and $\epsilon = 10^{-3}$ with $\rd=\rp$. 
  The min-max converse (Theorem \ref{thm:MC}) and the \gls{rcus} noncoherent bound (Theorem \ref{thm:RCUs_ML}) are included for reference.  }
 \label{fig:NN_comp}
 \end{figure}  

 In Fig. \ref{fig:NN_comp}, we compare the  \gls{pat}-\gls{rcus}-\gls{nn} achievability bound (Theorem~\ref{thm:RCUs_NN}) with its \gls{rce} relaxation (Corollary~\ref{thm:RCE_NN}) for the case $\rd=\rp$.
 We see that for $\ell = 28$, the gap between the bounds is about $0.5\dB$.

\subsection{Practical \gls{pat} Coding Schemes}

We discuss next the design of actual \gls{pat}-based coding schemes with moderate decoding complexity.
We shall focus for simplicity on the case $\ell=7$ and $\nc=24$.
Furthermore, we assume that $81$ information bits need to be transmitted in each codeword, which 
yields $R\approx 0.48$ bit/channel use.
We allocate $\np$ channel uses per coherence block to pilot symbols, and use
the remaining $\left(24-\np \right)$ channel uses to carry coded symbols belonging to a \gls{qpsk} 
constellation. 
Similar to~\cite{ostman17-02a}, we select a $\left(324, 81\right)$ binary quasi-cyclic code and puncture a suitable number of codeword 
bits to accommodate the pilot symbols within the prescribed $168$ channel uses. 
The code is obtained by tail-biting termination of a rate$-1/4$ nonsystematic convolutional code 
with memory~$14$ \cite[Table.~10.14]{johannesson_book}. 
The  minimum distance of the quasi-cyclic code is upper bounded by the free distance of the 
underlying convolution code, which is $36$.\footnote{This upper bound is expected to be tight 
because the ratio between the code dimension and the convolutional encoder memory is 
large~\cite{MaWolf1986}.} 
After encoding, a pseudo-random interleaving is applied to the codeword bits, followed by puncturing. 
For the chosen parameters, the number of punctured bits is $14 \np -12$  and the blocklength after puncturing (expressed this time in \emph{real} rather than \emph{complex} channel uses) is $336-14\np$.
At the receiver side, the pilot symbols are used to perform \gls{ml} channel estimation according to~\eqref{eq:ml_fading_estimation}. 
The bit-wise \gls{llr} are computed by assuming the estimates $\widehat{h}_k$, $k = 1,\dots,7$ to be perfect. 
Decoding is then performed via \gls{osd}~\cite{FosLin95}. 
The order of~\gls{osd} is set to $t=3$, which  provides a reasonable trade-off between performance and decoding complexity. 
The \gls{osd} builds a list $\mathcal{L}$ of $1+\sum_{i=1}^{t}$${81}\choose{i}$ $ = 88642$ channel input vectors corresponding to candidate codewords, out of which the decision is obtained as
\begin{equation}
\widehat{\vecx} = \argmax_{\vecx\in\mathcal{L}}
\prod_{k=1}^\ell \exp\lefto(-\vecnorm{\vecy^{(\text{d})}_k-\widehat{h}_k{\vecx}_k}^2\right)
\label{eq:OSD_dec}
\end{equation}
where $\vecx_k$ denote the vector of coded \gls{qpsk} symbols  transmitted over the $k$th coherence interval.
We shall refer to the decoder operating according to this rule as \gls{osd}--\gls{nn}. 
When the list $\mathcal{L}$ includes all input vectors corresponding to valid codewords, the decoding rule \eqref{eq:OSD_dec} is equivalent to SNN in~\eqref{eq:nn_metric}. 
We also analyze a second scheme, in which a re-estimation of the fading channel is performed by using the initial \gls{osd} decision $\widehat{\vecx}$. 
Specifically, $\widehat{\vecx}$ is used to update the \gls{ml} channel estimates, yielding new bit-wise \gls{llr}. 
A second \gls{osd} attempt is then performed with the updated input. 
We refer to this second scheme as \gls{osd} with re-estimation (\gls{osd}--REE).

In Fig. \ref{fig:codes}, we compare the performance of the \gls{osd}--\gls{nn} coding scheme to what predicted by the \gls{pat}-\gls{rcus}-\gls{nn} achievability bound (Theorem~\ref{thm:RCUs_NN}) for different values of $\np$, for the case $\rp=\rd$.
We see that the gap is within $1$ dB for all values of $\np$ considered here.
This shows that the performance reference provided by the \gls{pat}-\gls{rcus}-\gls{nn} achievability bound is accurate.
For the parameters considered in Fig. \ref{fig:codes}, setting $\np=4$ yields the best performance, as predicted by the \gls{pat}-\gls{rcus}-\gls{nn} bound.

In Fig. \ref{fig:codes_re}, we compare the  performance of the \gls{osd}--REE coding scheme with what predicted by the \gls{rcus}--\gls{pat}--\gls{ml} achievability bound in Theorem~\ref{thm:RCUs_Pilot}.
This bound is relevant since the \gls{osd}--REE coding scheme improves on the \gls{nn} decoding rule by allowing decision-driven channel re-estimation.
The gap between the bound and the code performance is now larger: about $1.3$ dB for $\epsilon=10^{-3}$ and $\np=4$.
This is due to the fact that the \gls{rcus}--\gls{pat}--\gls{ml} achievability bound assumes \gls{ml} decoding, which yield too optimistic performance estimates.
Comparing Figs.~\ref{fig:codes} and~\ref{fig:codes_re}, we see that the performance gains of the \gls{osd}--REE coding scheme over the \gls{osd}--\gls{nn} one are limited to  fractions of dBs, e.g., for $\np=4$ and $\epsilon= 10^{-3}$, the gain is about $0.5$ dB.

\begin{figure}[h]
\centering
 \begin{tikzpicture}
\pgfplotsset{
    scaled y ticks = false,
    width=\fwidth*0.35,
    height=\fwidth*0.3,
    axis on top,
    xmin=3,xmax=12,
    ymin = 1e-3,  ymax = 1e-1,
       axis x line*=bottom,
     xmode = linear,
     ymode = log,
       xlabel={$E\sub{b}/N_0$ [dB]},
         ylabel = $\epsilon$,
      	 xlabel near ticks,
     	 grid=both,
     	 title style={yshift=-6pt,}
}
    \begin{groupplot}[ 
        group style={
        group size=2 by 4,
        vertical sep=45pt,
        horizontal sep=35pt
        },
    ]
    \nextgroupplot[
            x filter/.code={\pgfmathparse{\pgfmathresult-10*log10(0.48)}},
            title={$\np=1$} ,
            legend style={at={($(0,0)+(1cm,1cm)$)},legend columns=4,fill=none,draw=black,anchor=center,align=center},
            legend to name=fred,
    ]

 \addplot [name path = MC, color = blue, solid,mark=*,mark options={solid, mark size=\marksize},line width=\linew] table [y index={0}, x index = {1}, col sep=comma] {./Data/bounds_error_vs_snr_NN.csv}coordinate[pos=0.4](p1);\addlegendentry{Min-max converse}
 \addplot [name path = DT, color = red, solid,mark=square*,mark options={solid, mark size=\marksize},line width=\linew] table [y index={0},x index = {2},  col sep=comma] {./Data/bounds_error_vs_snr_NN.csv}coordinate[pos=0.42](p1);\addlegendentry{\gls{rcus} noncoherent}
 \addplot [color=gray, opacity=0.2, forget plot] fill between[of=MC and DT];
 \addplot[solid,mark=triangle*,line width=\linew, color = black!30!green,mark options={solid, mark size=\marksize}] table [y index={0},x index = {3},  col sep=comma] {./Data/bounds_error_vs_snr_NN.csv}coordinate[pos=0.5](p1);\addlegendentry{\gls{rcus}--\gls{pat}--\gls{nn}}
 \addplot[solid,mark=diamond*,mark options={solid, mark size=\marksize},line width=\linew, color = orange] table [y index={1},x index = {0},  col sep=comma] {./Data/codes_np1.csv}coordinate[pos=0.5](p1);\addlegendentry{\gls{osd}-\gls{nn}}
    \coordinate (c1) at (rel axis cs:0,1);
    \nextgroupplot[   x filter/.code={\pgfmathparse{\pgfmathresult-10*log10(0.48)}},         title={$\np=2$}                ]

 \addplot [name path = MC, color = blue, solid,mark=*,mark options={solid, mark size=\marksize},line width=\linew] table [y index={0}, x index = {1}, col sep=comma] {./Data/bounds_error_vs_snr_NN.csv}coordinate[pos=0.4](p1);
 \addplot [name path = DT, color = red, solid,mark=square*,mark options={solid, mark size=\marksize},line width=\linew] table [y index={0},x index = {2},  col sep=comma] {./Data/bounds_error_vs_snr_NN.csv}coordinate[pos=0.42](p1);
 \addplot [color=gray, opacity=0.2, forget plot] fill between[of=MC and DT];
 \addplot [solid,mark=triangle*,mark options={solid, mark size=\marksize},line width=\linew, color =  black!30!green] table [y index={0},x index = {4},  col sep=comma] {./Data/bounds_error_vs_snr_NN.csv}coordinate[pos=0.5](p12);
 \addplot [solid,mark=diamond*,mark options={solid, mark size=\marksize},line width=\linew, color = orange] table [y index={1},x index = {0},  col sep=comma] {./Data/codes_np2.csv}coordinate[pos=0.5](p12);
    \coordinate (c2) at (rel axis cs:1,1);
    \nextgroupplot[ 	   x filter/.code={\pgfmathparse{\pgfmathresult-10*log10(0.48)}}  , title={$\np=4$}    ]

 \addplot [name path = MC, color = blue, solid,mark=*,mark options={solid, mark size=\marksize},line width=\linew] table [y index={0}, x index = {1}, col sep=comma] {./Data/bounds_error_vs_snr_NN.csv}coordinate[pos=0.4](p1);
 \addplot [name path = DT, color = red, solid,mark=square*,mark options={solid, mark size=\marksize},line width=\linew] table [y index={0},x index = {2},  col sep=comma] {./Data/bounds_error_vs_snr_NN.csv}coordinate[pos=0.42](p1);
 \addplot [color=gray, opacity=0.2, forget plot] fill between[of=MC and DT];
 \addplot [ solid,mark=triangle*,line width=\linew, color =  black!30!green,mark options={solid, mark size=\marksize}] table [y index={0},x index = {5},  col sep=comma] {./Data/bounds_error_vs_snr_NN.csv}coordinate[pos=0.5](p14);
 \addplot [ solid,mark=diamond*,mark options={solid, mark size=\marksize},line width=\linew, color =  orange] table [y index={1},x index = {0},  col sep=comma] {./Data/codes_np4.csv}coordinate[pos=0.5](p14);
    \nextgroupplot[              title={$\np=8$} ,x filter/.code={\pgfmathparse{\pgfmathresult-10*log10(0.48)}}    ]          
 \addplot [name path = MC, color = blue, solid,mark=*,mark options={solid, mark size=\marksize},line width=\linew] table [y index={0}, x index = {1}, col sep=comma] {./Data/bounds_error_vs_snr_NN.csv}coordinate[pos=0.4](p1);
 \addplot [name path = DT, color = red, solid,mark=square*,mark options={solid, mark size=\marksize},line width=\linew] table [y index={0},x index = {2},  col sep=comma] {./Data/bounds_error_vs_snr_NN.csv}coordinate[pos=0.42](p1);
 \addplot [color=gray, opacity=0.2, forget plot] fill between[of=MC and DT];
 \addplot [solid,mark=triangle*,line width=\linew, color =  black!30!green,mark options={solid, mark size=\marksize}] table [y index={0},x index = {7},  col sep=comma] {./Data/bounds_error_vs_snr_NN.csv}coordinate[pos=0.7](p18);
 \addplot [solid,mark=diamond*, line width=\linew, color = orange,mark options={solid, mark size=\marksize}] table [y index={1},x index = {0},  col sep=comma] {./Data/codes_np8.csv}coordinate[pos=0.7](p18);
    \end{groupplot}
    \coordinate (c3) at ($(c1)!.5!(c2)$);
    \node[below] at (c3 |- current bounding box.south)
      {\pgfplotslegendfromname{fred}};
\end{tikzpicture}%
  \caption{Performance of the \gls{osd}--\gls{nn} coding scheme for $\np=\lrbo{1,2,4,8}$; the \gls{rcus}--\gls{pat}--\gls{nn} (Theorem \ref{thm:RCUs_NN}), the min-max converse (Theorem \ref{thm:MC}), and the \gls{rcus} noncoherent bound (Theorem \ref{thm:RCUs_ML}) are also plotted for reference; $\nc=24$, $\ell=7$, $R=0.48$ bit/channel use, and $\kappa=0$.}
 \label{fig:codes}
 \end{figure} 
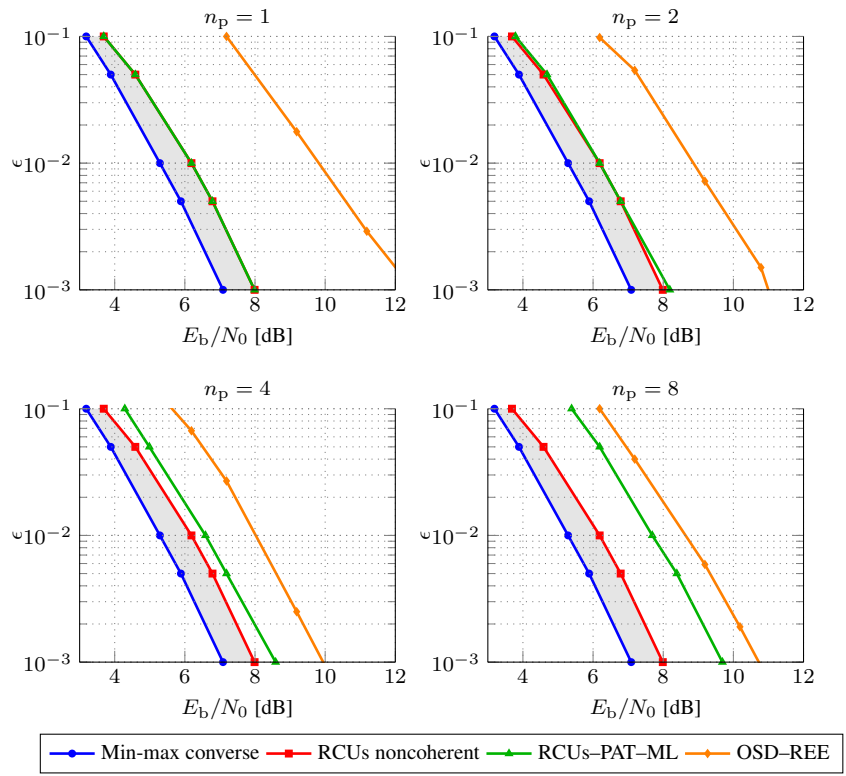
\begin{figure}[h]
\centering
 \begin{tikzpicture}
\pgfplotsset{
    scaled y ticks = false,
    width=\fwidth*0.35,
    height=\fwidth*0.3,
    axis on top,
    xmin=3,xmax=12,
    ymin = 1e-3,  ymax = 1e-1,
       axis x line*=bottom,
     xmode = linear,
     ymode = log,
       xlabel={$E\sub{b}/N_0$ [dB]},
         ylabel = $\epsilon$,
      	 xlabel near ticks,
     	 grid=both,
     	 title style={yshift=-6pt,}
}
    \begin{groupplot}[ 
        group style={
        group size=2 by 4,
        vertical sep=45pt,
        horizontal sep=35pt
        },
    ]
    \nextgroupplot[
            x filter/.code={\pgfmathparse{\pgfmathresult-10*log10(0.48)}},
            title={$\np=1$} ,
            legend style={at={($(0,0)+(1cm,1cm)$)},legend columns=4,fill=none,draw=black,anchor=center,align=center},
            legend to name=fred,
    ]
 \addplot [name path = MC, color = blue, solid,mark=*,mark options={solid, mark size=\marksize},line width=\linew] table [y index={0}, x index = {1}, col sep=comma] {./Data/bounds_error_vs_snr_ML.csv}coordinate[pos=0.4](p1);\addlegendentry{Min-max converse}
 \addplot [name path = DT, color = red, solid,mark=square*,mark options={solid, mark size=\marksize},line width=\linew] table [y index={0},x index = {2},  col sep=comma] {./Data/bounds_error_vs_snr_ML.csv}coordinate[pos=0.42](p1);\addlegendentry{\gls{rcus} noncoherent}
 \addplot [color=gray, opacity=0.2, forget plot] fill between[of=MC and DT];
 \addplot[solid,mark=triangle*,line width=\linew, color = black!30!green,mark options={solid, mark size=\marksize}] table [y index={0},x index = {2},  col sep=comma] {./Data/bounds_error_vs_snr_ML.csv}coordinate[pos=0.5](p1);\addlegendentry{\gls{rcus}--\gls{pat}--\gls{ml}}
 \addplot[solid,mark=diamond*,line width=\linew, color = orange,mark options={solid, mark size=\marksize}] table [y index={1},x index = {0},  col sep=comma] {./Data/codes_np1re.csv}coordinate[pos=0.5](p1);\addlegendentry{\gls{osd}--REE}

    \coordinate (c1) at (rel axis cs:0,1);
    \nextgroupplot[   x filter/.code={\pgfmathparse{\pgfmathresult-10*log10(0.48)}},         title={$\np=2$}                ]
 \addplot [name path = MC, color = blue, solid,mark=*,mark options={solid, mark size=\marksize},line width=\linew] table [y index={0}, x index = {1}, col sep=comma] {./Data/bounds_error_vs_snr_ML.csv}coordinate[pos=0.4](p1);
 \addplot [name path = DT, color = red, solid,mark=square*,mark options={solid, mark size=\marksize},line width=\linew] table [y index={0},x index = {2},  col sep=comma] {./Data/bounds_error_vs_snr_ML.csv}coordinate[pos=0.42](p1);
 \addplot [color=gray, opacity=0.2, forget plot] fill between[of=MC and DT];
 \addplot [ solid,mark=triangle*,mark options={solid, mark size=\marksize},line width=\linew, color = black!30!green] table [y index={0},x index = {4},  col sep=comma] {./Data/bounds_error_vs_snr_ML.csv}coordinate[pos=0.5](p14);
 \addplot [ solid,mark=diamond*,mark options={solid, mark size=\marksize},line width=\linew, color = orange] table [y index={1},x index = {0},  col sep=comma] {./Data/codes_np2re.csv}coordinate[pos=0.5](p14);
    \coordinate (c2) at (rel axis cs:1,1);
    \nextgroupplot[ 	   x filter/.code={\pgfmathparse{\pgfmathresult-10*log10(0.48)}}  , title={$\np=4$}    ]
 \addplot [name path = MC, color = blue, solid,mark=*,mark options={solid, mark size=\marksize},line width=\linew] table [y index={0}, x index = {1}, col sep=comma] {./Data/bounds_error_vs_snr_ML.csv}coordinate[pos=0.4](p1);
 \addplot [name path = DT, color = red, solid,mark=square*,mark options={solid, mark size=\marksize},line width=\linew] table [y index={0},x index = {2},  col sep=comma] {./Data/bounds_error_vs_snr_ML.csv}coordinate[pos=0.42](p1);
 \addplot [color=gray, opacity=0.2, forget plot] fill between[of=MC and DT];
 \addplot [solid,mark=triangle*,mark options={solid, mark size=\marksize},line width=\linew, color = black!30!green] table [y index={0},x index = {5},  col sep=comma] {./Data/bounds_error_vs_snr_ML.csv}coordinate[pos=0.6](p16);
 \addplot [solid,mark=diamond*,mark options={solid, mark size=\marksize},line width=\linew, color = orange] table [y index={1},x index = {0},  col sep=comma] {./Data/codes_np4re.csv}coordinate[pos=0.6](p16);

    \nextgroupplot[              title={$\np=8$} ,x filter/.code={\pgfmathparse{\pgfmathresult-10*log10(0.48)}}    ]          
       \addplot [name path = MC, color = blue, solid,mark=*,mark options={solid, mark size=\marksize},line width=\linew] table [y index={0}, x index = {1}, col sep=comma] {./Data/bounds_error_vs_snr_ML.csv}coordinate[pos=0.4](p1);
	 \addplot [name path = DT, color = red, solid,mark=square*,mark options={solid, mark size=\marksize},line width=\linew] table [y index={0},x index = {2},  col sep=comma] {./Data/bounds_error_vs_snr_ML.csv}coordinate[pos=0.42](p1);
	 \addplot [color=gray, opacity=0.2, forget plot] fill between[of=MC and DT];
 \addplot [solid,mark=triangle*,mark options={solid, mark size=\marksize},line width=\linew, color = black!30!green] table [y index={0},x index = {7},  col sep=comma] {./Data/bounds_error_vs_snr_ML.csv}coordinate[pos=0.7](p18);
 \addplot [solid,mark=diamond*,mark options={solid, mark size=\marksize},line width=\linew, color = orange] table [y index={1},x index = {0},  col sep=comma] {./Data/codes_np8re.csv}coordinate[pos=0.7](p18);

    \end{groupplot}
    \coordinate (c3) at ($(c1)!.5!(c2)$);
    \node[below] at (c3 |- current bounding box.south)
      {\pgfplotslegendfromname{fred}};
\end{tikzpicture}%
\caption{Performance of the \gls{osd}--REE coding scheme for $\np=\lrbo{1,2,4,8}$; the  \gls{rcus}--\gls{pat}--\gls{ml}  bound (Theorem \ref{thm:RCUs_Pilot}), the min-max converse (Theorem \ref{thm:MC}), and the \gls{rcus} noncoherent bound (Theorem \ref{thm:RCUs_ML}) are  plotted for reference;  $\nc=24$, $\ell=7$, and $R=0.48$ bit per channel use, and $\kappa=0$.}
 \label{fig:codes_re}
 \end{figure}

\section{Conclusion}
We presented bounds on the maximum coding rate achievable over a \gls{siso} Rician memoryless block-fading channel under the assumption of no \emph{a priori} \gls{csi}.
Specifically, we presented converse and achievability bounds on the maximum coding rate that generalize and tighten the bounds previously reported in~\cite{durisi16-02a,Ostman17-03a}.
Our two achievability bounds, built upon the \gls{rcus} bound, allow one to compare the performance of noncoherent and \gls{pat} schemes.
As in~\cite{durisi16-02a,Ostman17-03a} our converse bound relies on the min-max converse. 

Through a numerical investigation, we showed that our converse and achievability bounds delimit tightly the maximum coding rate, for a large range of SNR and Rician $\kappa$-factor values, and allow one to identify---for given coding rate and packet size---the optimum number of coherence blocks to code over in order to minimize the energy per bit required to attain a target packet error probability.

Furthermore, our achievability bounds reveal that noncoherent transmission is more energy efficient than \gls{pat} even when the number of pilot symbols and their power is optimized.\footnote{We limit our comparison to the two achievability bounds because no tight converse bound for the \gls{pat} case is available, even asymptotically.} 
When the power of the pilot symbols is optimized, one pilot symbol per coherence block turns out to suffice---a nonasymptotic counterpart of the result obtained in~\cite{hassibi03-04a}. 

We finally designed an actual \gls{pat} scheme based on punctured tail-biting quasi-cyclic codes and a decoder that, using \gls{osd}, performs \gls{nn} detection based on \gls{ml} channel estimates.
A comparison between the \gls{pat} scheme and our bounds reveals that the bounds provide accurate guidelines on the design of actual \gls{pat} schemes.
We also discussed how the performance of the decoder can be further improved (without hampering its relatively low computational complexity) by accounting for the inaccuracy of the channel estimates.

An important final remark is that our comparison between noncoherent and PAT schemes is somewhat biased towards the noncoherent case.
Indeed, our \gls{rcus} noncoherent bound relies on \gls{ml} decoding (which implies also knowledge of the fading law), whereas both \gls{rcus}--\gls{pat}--\gls{nn} and \gls{osd}--\gls{nn} rely on a lower-complexity \gls{nn} decoder and require no knowledge of the fading law.
Designing low-complexity noncoherent coding schemes able to approach our \gls{rcus} noncoherent bound is an important open issue.



\appendix[] \label{app:A}
\subsection{Auxiliary Lemmas} 
\label{sec:auxiliary_lemmas}

We state next  two  lemmas that will be useful for proving our achievability and converse bounds on $R^*$.

\begin{lem}
\label{lem:ustm_induced_conditioned_pdf}
Let $\randvecx$ be an isotropically distributed vector in $\complexset^{\nc}$  with norm equal to $\sqrt{\rho \nc}$, let $H\sim \cgauss{\mu\sub{H}}{\sigma\sub{H}^2}$, and let $\randvecw\sim \cgauss{0}{\sigma\sub{w}^2\matI_{\nc}}$
Furthermore, let $\randvecy = H\randvecx  + \randvecw$. 
The conditional \gls{pdf} of $\randvecy$ given  $H=h$ is 
\bie
\label{eq:shell_induced_conditioned_out_pdf}
  P_{\randvecy\given H}\lro{\vecy\given h} &=& \frac{\Gamma\lro{\nc}\exp\lro{-\frac{\vecnorm{\vecy}^2+\abs{h}^2 \rho \nc}{\sigma\sub{w}^2}} 
  }{\pi^{\nc} \sigma\sub{w}^2 \lro{\vecnorm{\vecy}\abs{h} \sqrt{\rho \nc}}^{\nc-1}}\,  I_{\nc-1}\lro{ \frac{2\vecnorm{\vecy}\abs{h} \sqrt{\rho \nc}}{\sigma^2\sub{w}}}.
\eie
\end{lem}
\begin{IEEEproof}
 Under the assumptions of Lemma~\ref{lem:ustm_induced_conditioned_pdf},  the random variable $(\sigma\sub{w}^2/2)\vecnorm{\vecy}^2$  follows (given~$h$) a noncentral $\chi$-squared distribution with $2\nc$ degrees of freedom and noncentrality parameter $2\abs{h}^2\nc\rho/\sigma\sub{w}^2$.
  Furthermore, the output vector $\vecy$ is isotropically distributed.
  We then obtain~\eqref{eq:shell_induced_conditioned_out_pdf} by recalling that the surface area of an $\nc$-dimensional complex sphere of radius $\sqrt{\nc\rho}$ is
  \begin{equation}\label{eq:radius}
    \frac{2\pi^{\nc}(\sqrt{\nc\rho})^{2\nc-1}}{\Gamma(\nc)}.
  \end{equation}
  %
\end{IEEEproof}

\begin{lem}
\label{lem:ustm_induced_pdf}
 Under the assumptions of Lemma~\ref{lem:ustm_induced_conditioned_pdf}, the \gls{pdf} of $\randvecy$ is 
\bie
\label{eq:shell_induced_out_pdf}
  P_{\randvecy}\lro{\vecy} &=& \frac{\Gamma\lro{\nc}
  \exp\lro{-\frac{\vecnorm{\vecy}^2}{\sigma\sub{w}^2}-\frac{\abs{\mu\sub{H} }^2}{\sigma\sub{H}^2}} 
  }
  {\pi^{\nc}\sigma\sub{w}^2 \sigma\sub{H}^2} 
 \int_{0}^{\infty} \frac{\exp\lro{-z\lro{\frac{\rho\nc}{\sigma\sub{w}^{2}} + \frac{1}{\sigma\sub{H}^{2}}} }}{ \lro{\vecnorm{\vecy}\sqrt{\rho \nc z}}^{\nc-1} } \nonumber \\
&  &\times  I_{\nc-1}\lro{ \frac{2\vecnorm{\vecy} \sqrt{\rho \nc z}}{\sigma^2\sub{w}}} 
I_0\lro{\frac{2\abs{\mu\sub{H}}\sqrt{z}}{\sigma\sub{H}^{2}}} \mathrm{d}z.
\eie
\end{lem}
\begin{IEEEproof}
We obtain~\eqref{eq:shell_induced_out_pdf} by averaging~\eqref{eq:shell_induced_conditioned_out_pdf} over $\abs{H}^2$, which has \gls{pdf}
 \begin{IEEEeqnarray}{rCl}\label{eq:pdf_of_channel_gain}
 P_{\abs{H}^2}\lro{z} = \frac{1}{\sigma\sub{H}^2} \exp\lro{-\frac{1}{\sigma\sub{H}^2}\lro{z + \abs{\mu\sub{H}}^2}} I_0\lro{\frac{2\abs{\mu\sub{H}}\sqrt{z}}{\sigma\sub{H}^{2}}}.
 \end{IEEEeqnarray}
 \end{IEEEproof}
%


\subsection{Proof of Theorem \ref{thm:RCUs_ML}} 
 \label{app:RCUs_proof}
We let $\randvecx_k = \sqrt{\nc \rho} \randvecu_k$ where $\lrb{\randvecu_k}_{k=1}^\ell$ are independent and isotropically distributed unitary vectors in $\complexset^{\nc}$.
For the chosen decoding metric~\eqref{eq:ML_decoder}, the generalized information density in~\eqref{eq:generalized_inf_dens} can be decomposed as
\bie
\label{eq:info_dens}
i^{\ell}_s\lro{\vecu^{\ell}, \vecy^{\ell}} &=& \sum_{k=1}^\ell i_s\lro{\vecu_k, \vecy_k} =
 \sum_{k=1}^\ell \log\frac{P_{\randvecy \given \randvecu}\lro{\vecy_k \given \vecu_k}^s}{ \Ex{}{P_{\randvecy\given \randvecu}\lro{\vecy_k\given \randvecu_k}^s}}
\eie
where
\bie
\label{eq:cond_distr_unit}
  P_{\randvecy\given \randvecu=\vecu_k}=\cgauss{ \mu\sub{H} \sqrt{\nc\rho} \vecu_k}{ \matSigma_k }
\eie%
with $\matSigma_k=\matI_{\nc}+\sigma\sub{H}^2\nc\rho \vecu_k \herm{\vecu}_k$.
To evaluate the expected value in~\eqref{eq:info_dens}, it is convenient to express $P_{\randvecy \given \randvecu}\lro{\vecy_k \given \vecu_k} ^s$ as a scalar times a Gaussian \gls{pdf} as follows:
\begin{IEEEeqnarray}{rCl} 
\label{eq:rewritten_gaussian}
P_{\randvecy \given \randvecu}\lro{\vecy_k \given \vecu_k} ^s &=& \lro{\pi^{\nc} \det\lro{\matSigma_k}}^{1-s} s^{-\nc} P_{\widetilde{\randvecy} \given \randvecu}\lro{\vecy_k \given \vecu_k} \\
&=&
\lro{\pi^{\nc} \lro{1+\rho\nc\sigma^2\sub{H} }}^{1-s} s^{-\nc} P_{\widetilde{\randvecy} \given \randvecu}\lro{\vecy_k \given \vecu_k} \label{eq:rewritten_gaussian_v2}
\end{IEEEeqnarray}
where $P_{\widetilde{\randvecy} \given \randvecu=\vecu_k} = \cgauss{\mu\sub{H} \sqrt{\nc\rho} \vecu_k}{s^{-1}\matSigma_k}$.
Note now that the conditional \gls{pdf} $P_{\widetilde{\randvecy} \given \randvecu}$ describes a channel with  input-output relation $\widetilde{\randvecy} = \sqrt{\nc\snr}\widetilde{H} \randvecu + \widetilde{\randvecw}$, where $\randvecu$ is an $\nc$-dimensional isotropically distributed unitary vector,  $\widetilde{H} \sim \cgauss{\mu\sub{H}}{ s^{-1}\sigma\sub{H}^2}$, and $\widetilde{\randvecw} \sim \cgauss{0}{s^{-1}\matI_{\nc}}$.
Applying  Lemma~\ref{lem:ustm_induced_pdf} in Appendix~\ref{app:A} to this channel (which entails replacing $\sigma\sub{H}^2$ in~\eqref{eq:shell_induced_out_pdf} by $s^{-1}\sigma\sub{H}^2$ and $\sigma\sub{w}^2$ by $s^{-1}$) we conclude that
\begin{multline}
\Ex{}{P_{\widetilde{\randvecy}\given \randvecu}\lro{\vecy_k\given \randvecu_k}} = \frac{\Gamma\lro{\nc} s^{2}\exp\lro{-s\vecnorm{\vecy_k}^2-s\frac{\abs{\mu\sub{H}}^2}{\sigma\sub{H}^{2}}} }{\pi^{\nc}  \sigma\sub{H}^2} \\
\times \int_{0}^{\infty}\frac{\exp\lro{-s\lro{\rho\nc + \frac{1}{\sigma\sub{H}^{2}}}z} }{\lro{ \vecnorm{\vecy_k}\sqrt{\rho\nc z}}^{\nc-1}} I_{\nc-1}\lro{2s\vecnorm{\vecy_k}\sqrt{\rho\nc z }} I_0\lro{\frac{2s\abs{\mu\sub{H}} \sqrt{z}}{\sigma\sub{H}^{2}}}\mathrm{d}z.
\end{multline}
It follows then from~\eqref{eq:rewritten_gaussian_v2} that
\begin{multline}\label{eq:output_dist_s}
\Ex{}{P_{\randvecy\given \randvecu}\lro{\vecy_k\given \randvecu_k}^s} = \frac{\Gamma\lro{\nc} s^{2-\nc}\exp\lro{-s\vecnorm{\vecy_k}^2-s\frac{\abs{\mu\sub{H}}^2}{\sigma\sub{H}^{2}}} }{\pi^{s\nc}  \lro{1+\rho\nc\sigma^2\sub{H}}^{s-1} \sigma\sub{H}^2} \\
\times \int_{0}^{\infty}\frac{\exp\lro{-s\lro{\rho\nc + \frac{1}{\sigma\sub{H}^{2}}}z} }{\lro{ \vecnorm{\vecy_k}\sqrt{\rho\nc z}}^{\nc-1}} I_{\nc-1}\lro{2s\vecnorm{\vecy_k}\sqrt{\rho\nc z }} I_0\lro{\frac{2s\abs{\mu\sub{H}} \sqrt{z}}{\sigma\sub{H}^{2}}}\mathrm{d}z.
\end{multline}

Finally, to evaluate the expectation in the \gls{rcus} bound~\eqref{eq:rcus-original}, we observe that~\eqref{eq:cond_distr_unit} and~\eqref{eq:output_dist_s} imply that for every $\nc\times\nc$ unitary matrix $\matV$,
\begin{equation}\label{eq:trasformation_generalized_inf_dens}
  i_s\lro{\herm{\matV} \vecu_k, \vecy_k} = i_s\lro{\vecu_k, \matV\vecy_k}.
\end{equation}
This in turn implies that when $\randvecy_k \sim P_{\randvecy \given \randvecu=\vecu_k}$
the probability distribution of $i_s\lro{\vecu_k, \randvecy_k}$
does not depend on $\vecu_k$.
Hence, we can set without loss of generality $\vecu_k = \tp{\lrho{1,0,\dots,0}}$, $k = 1,\dots, \ell$.
For this choice of $\{\vecu_k\}$, it  follows from~\eqref{eq:cond_distr_unit} and~\eqref{eq:output_dist_s} that $i_s\lro{\vecu_k, \randvecy_k}$ has the same distribution as the random variable $S^s_k$ defined in~\eqref{eq:Sk}.
%

\subsection{Proof of Corollary \ref{thm:RCE_ML}} 
 \label{app:RCE_ML_proof}
We evaluate Corollary~\ref{col:grce} for $\randvecx=\sqrt{\nc\snr}\randvecu$ where $\randvecu$ is  unitary and isotropically distributed. 
Furthermore, we choose the \gls{ml} decoding metric~\eqref{eq:ML_decoder}. 
For this choice, the maximum over $s$ in the Gallager's function for mismatch decoding~\eqref{eq:generalized_random_coding_error_exponent} is achieved by $s=1/(1+\tau)$~\cite[p.~137]{gallager68a}.
Let now   $F_0\lro{\tau} = e^{-E_0\lro{\tau,(1+\tau)^{-1}}}$, where $E_0\lro{\tau, (1+\tau)^{-1}}$ is defined in~\eqref{eq:gallager function for mismatch decoding}. 
Standard manipulations of the generalized information density reveal that 
%
\begin{IEEEeqnarray}{rCl}
\label{eq:MF01}
F_0\lro{\tau} = \int_{\complexset^{\nc}}\Ex{}{ P_{\randvecy\given \randvecu}\lro{\vecy  \given \randvecu}^{\frac{1}{1+\tau}} }^{1+\tau} d\vecy.
\end{IEEEeqnarray}
Note now that the expectation inside the integral in~\eqref{eq:MF01} can be computed  as in Appendix~\ref{app:RCUs_proof}; specifically, its value coincides with the right-hand side of~\eqref{eq:output_dist_s} provided that one replaces $s$ in~\eqref{eq:output_dist_s} with $(1+\tau)^{-1}$.
Substituting this expression in~\eqref{eq:MF01} and computing the integral in spherical coordinate, we obtain~\eqref{eq:gallager_function_noncoherent}.

\subsection{Proof of Theorem \ref{thm:RCUs_NN}} 
 \label{app:RCUs_NN_proof}
 We use the \gls{pat} scheme described in Section~\ref{sec:pilots_NN}.
We let $\randvecx_k^{\lro{\text{d}}} = \sqrt{ \rd\nd} \randvecu_k^{\lro{\text{d}}}$ where $\bigl\{\randvecu_k^{\lro{\text{d}}}\bigr\}_{k=1}^\ell$ are $\nd$-dimensional independent and isotropically distributed unitary vectors. 
The pilot symbols and the corresponding $\np$-dimensional received vectors are used to obtain a \gls{ml} estimate of the fading according to~\eqref{eq:ml_fading_estimation}.
We assume that the receiver uses the decoding \gls{nn} decoding metric~\eqref{eq:nn_metric}.
A decoder that operates according to~\eqref{eq:nn_metric} treats the channel estimates $\widehat{h}_k$ as perfect, which is equivalent to assuming that 
\begin{IEEEeqnarray}{rCl} \label{eq:cond_nn_pdf}
	\randvecy^{\lro{\text{d}}}_k\distas {P}_{\widetilde{\randvecy}^{\lro{\text{d}}} \given  \widehat{H} =\widehat{h}_k, \randvecu^{\lro{\text{d}}}=\vecu^{\lro{\text{d}}}_k} = \cgauss{\widehat{h}_k\sqrt{\rd\nd}\vecu^{\lro{\text{d}}}_k}{\matI_{\nd}}.
\end{IEEEeqnarray}
This allows us to rewrite the generalized information density in~\eqref{eq:gen_inf_dens_block_additive} as
\bie
\label{eq:info_dens_NN}
 i^{\ell}_s\lro{\vecx^{\ell}, \vecy^{\ell}} = \sum_{k=1}^\ell i_s\lro{\vecu^{\lro{\text{d}}}_k, \vecy^{\lro{\text{d}}}_k, \widehat{h}_k} =
 \sum_{k=1}^\ell \log\frac{ {P}_{\widetilde{\randvecy}^{\lro{\text{d}}} \given  \widehat{H},\randvecu^{\lro{\text{d}}} } \lro{\vecy^{\lro{\text{d}}}_k \given  \widehat{h}_k, \vecu^{\lro{\text{d}}}_k}^s}{ \Ex{}{ {P}_{\widetilde{\randvecy}^{\lro{\text{d}}} \given  \widehat{H}, \randvecu^{\lro{\text{d}}} } \lro{\vecy^{\lro{\text{d}}}_k\given  \widehat{h}_k, \randvecu^{\lro{\text{d}}}_k}^s}}.
\eie
To evaluate the expected value in~\eqref{eq:info_dens_NN}, we proceed similarly as in Appendix~\ref{app:RCUs_proof} and obtain
%
%
\begin{multline}
\Ex{}{ P_{\widetilde{\randvecy}^{\lro{\text{d}}} \given \widehat{H},\randvecu^{\lro{\text{d}}} } \lro{\vecy_k^{\lro{\text{d}}}\given  \widehat{h}_k, \randvecu^{\lro{\text{d}}}_k}^s} \\
=\frac{\Gamma\lro{\nd} \exp\lro{-s\lro{\vecnorm{\vecy_k}^2+\rd\nd\abs{\widehat{h}_k}^2}} }{\pi^{s\nd}\lro{s\vecnorm{\vecy_k} \abs{\widehat{h}_k} \sqrt{\rd\nd }}^{\nd-1}}   I_{\nd-1}\lro{2s\vecnorm{\vecy_k}  \abs{\widehat{h}_k}\sqrt{\rd\nd }}.\label{eq:nn_output_pdf}
\end{multline}

Finally, to evaluate the expectation in the \gls{rcus} bound~\eqref{eq:rcus-original}, we observe that~\eqref{eq:cond_nn_pdf} and~\eqref{eq:nn_output_pdf} imply that for every $\nc\times\nc$ unitary matrix $\matV$,
\bie
i_s\lro{\herm{\matV} \vecu_k^{\lro{\text{d}}}, \vecy^{\lro{\text{d}}}_k, \widehat{H}_k} = i_s\lro{\vecu_k^{\lro{\text{d}}}, \matV\vecy^{\lro{\text{d}}}_k, \widehat{H}_k}.
\eie
This in turn implies that when $\randvecy^{\lro{\text{d}}} \sim P_{\randvecy^{\lro{\text{d}}} \given H=h_k, \randvecu^{\lro{\text{d}}}=\vecu_k^{\lro{\text{d}}}}$ (the actual conditional \gls{pdf} of the output vector), the probability distribution of $i_s(\vecu_k^{\lro{\text{d}}},\randvecy^{\lro{\text{d}}}_k, \widehat{H}_k)$ does not depend on $\vecu_k^{\lro{\text{d}}}$.
Hence, we can set, without loss of generality, $\vecu_k^{\lro{\text{d}}} = \tp{\lrho{1,0,\dots,0}}$, $k = 1,\dots, \ell$.
One can finally show that under this choice of input vector,  $i_s(\vecu_k^{\lro{\text{d}}}, \randvecy^{\lro{\text{d}}}_k, \widehat{H}_k)$ has the same distribution as the random variable $T^s_k$ in \eqref{eq:Tk}.

\subsection{Proof of Corollary \ref{thm:RCE_NN}} 
 \label{app:RCE_NN_proof}
 We use the \gls{pat} scheme introduced in Section~\ref{sec:pilots_NN} and evaluate Corollary~\ref{col:grce} for $\randvecx^{\lro{\text{d}}}=\sqrt{\nc\snr}\randvecu^{\lro{\text{d}}}$ where $\randvecu^{\lro{\text{d}}}$ is an $\nd$-dimensional unitary and isotropically distributed random vector.\footnote{To keep the notation compact, we shall denote $\randvecu^{\lro{\text{d}}}$ and the corresponding output vector $\randvecy^{\lro{\text{d}}}$ simply as $\randvecu$ and $\randvecy$.} 
Furthermore, we choose the \gls{nn} decoding metric~\eqref{eq:nn_metric}.
Assume that \gls{ml} channel estimation yields the channel estimate $\widehat{H}=\widehat{h}$. 
Let $F_0\lro{\tau,s,\widehat{h}}=\exp(-E_0(\tau,s,\widehat{h}))$, where  $E_0(\tau,s,\widehat{h})$ is defined as in~\eqref{eq:gallager function for mismatch decoding} (we indicate explicitly its dependency from the channel estimate $\widehat{h}$).
Furthermore, let 
\begin{equation}\label{eq:snn_distribution}
    {P}_{\widetilde{\randvecy} \given \randvecu=\vecu,\widehat{H} =\widehat{h}} = \cgauss{\widehat{h}\sqrt{\rd\nd}\vecu}{\matI_{\nd}}.
\end{equation}
Our assumptions imply that
\begin{equation}\label{eq:F0_function_expression}
  F_0\lro{\tau,s,\widehat{h}} =\Ex{}{\Ex{\randvecu'}{\left(\frac{{P}_{\widetilde{\randvecy} \given  \randvecu,\widehat{H}}(\randvecy\given\randvecu',\widehat{h})}{{P}_{\widetilde{\randvecy} \given   \randvecu, \widehat{H}}(\randvecy\given\randvecu,\widehat{h})}\right)^s \middle | \randvecu,\randvecy}^\tau }
\end{equation}
where $P_{\randvecy,\randvecu,\randvecu'}(\vecy,\vecu,\vecu')=P_{\randvecu}(\vecu')P_{\randvecu}(\vecu')P_{\randvecy\given\randvecu,\widehat{H}}(\vecy|\vecu,\widehat{h})$. 
Here, $P_{\randvecy\given\randvecu,\widehat{H}}$ is the conditional output distribution of the channel, given the input $\vecu$ and the channel estimate $\widehat{h}$.
Since $P_{H\given \widehat{H}=\widehat{h}} = \cgauss{\mu\sub{p}(\widehat{h})}{\sigma\sub{p}^2}$ where $\mu\sub{p}(\widehat{h})$ and $\sigma\sub{p}^2$ are defined in~\eqref{eq:parameters_eq_channel}, we conclude that
\begin{equation}\label{eq:conditional_output_dist_given_estimate}
  P_{\randvecy\given \randvecu,\widehat{H}=\widehat{h}} = \cgauss{\sqrt{\rd\nd} \mu\sub{p}(\widehat{h}) \vecu}{\rd\nd\sigma\sub{p}^2 \vecu\herm{\vecu}+\matI_{\nd}}.
\end{equation}

We next evaluate the two expectations in~\eqref{eq:F0_function_expression}.
Using~\eqref{eq:snn_distribution} and~\eqref{eq:nn_output_pdf}, we can write the inner expectation as 
\begin{multline}
    \Ex{\randvecu'}{\left(\frac{{P}_{\widetilde{\randvecy} \given  \randvecu,\widehat{H}}(\vecy\given\randvecu',\widehat{h})}{{P}_{\widetilde{\randvecy} \given  \randvecu, \widehat{H},}(\vecy\given\vecu,\widehat{h})}\right)^s} 
     \\
    =\frac{\Gamma\lro{\nd} \exp\lro{s\lro{\vecnorm{\vecy - \sqrt{ \rd\nd}\vecu\widehat{h}}^2-\vecnorm{\vecy}^2-\rd\nd\abs{\widehat{h}}^2}}  }{(s \vecnorm{\vecy}\abs{\widehat{h}} \sqrt{\rd\nd })^{\nd-1}}  I_{\nd-1}(2s\vecnorm{\vecy}  \abs{\widehat{h}}\sqrt{\rd\nd }). \label{eq:inner_expectation_RCE} 
\end{multline}
Substituting~\eqref{eq:inner_expectation_RCE} into~\eqref{eq:F0_function_expression} and using~\eqref{eq:conditional_output_dist_given_estimate}, we obtain
\begin{IEEEeqnarray}{rCl} 
F_0\lro{\tau,s,\widehat{h}}
&=&
\int_{\complexset^{\nd}}  \frac{\Gamma\lro{\nd}^{\tau}\exp\lro{\frac{\rd \nd}{u} \lro{\abs{a(\widehat{h})}^2  - \abs{\mu\sub{p}(\widehat{h})}^2} }}{\pi^{\nd} \lro{1+\sigma\sub{p}^2\ \rd\nd}  (s\abs{\widehat{h}}\sqrt{\vecnorm{\vecy}^2 \rd\nd })^{\tau\lro{\nd-1}}} I_{\nd-1}\bigl(2s\abs{\widehat{h}}\sqrt{\vecnorm{\vecy}^2 \rd\nd }\bigr)^\tau  \nonumber \\
&&
 \times \Ex{\randvecu}{  e^{-\herm{\lro{\vecy - \sqrt{ \rd\nd} a(\widehat{h})\randvecu}} \lro{ \rd\nd \sigma\sub{p}^2 \randvecu\herm{\randvecu} +  \matI_{\nd} }^{-1}\lro{\vecy - \sqrt{ \rd\nd} a(\widehat{h})\randvecu} }}  \mathrm{d}\vecy \label{eq:RCE_NN_proof_eq1}
\end{IEEEeqnarray}
where $a(\widehat{h}) = \mu\sub{p}(\widehat{h}) - \widehat{h}s\tau u$ and $u = 1 + \sigma\sub{p}^2 \rd \nd$. 
Note that the term inside the expectation is proportional to the law of a channel with input-output relation $\widetilde{\randvecy} = \sqrt{ \rd\nd}\widetilde{H} \randvecu + \randvecw$, where $\widetilde{H} \sim \cgauss{a(\widehat{h})}{\sigma\sub{p}^2}$ and $\randvecw \sim \cgauss{0}{\matI_{\nd}}$.
Using Lemma \ref{lem:ustm_induced_pdf} in Appendix \ref{app:A} to evaluate this expectation, and computing the outer integral in spherical coordinates, we obtain
 \begin{IEEEeqnarray}{rCl}\label{eq:f0_final}
F_0\lro{\tau,s,\widehat{h}} &=&\Gamma\lro{\nd} ^\tau \sigma\sub{p}^{-2}   \exp\lro{\abs{a(\widehat{h})}^2 \lro{\frac{\rd \nd}{u}  - \frac{1}{\sigma\sub{p}^2}} - \frac{\abs{\mu\sub{p}(\widehat{h})}^2 \rd\nd}{u}}  \nonumber\\
&&\times  \int_{0}^{\infty} \frac{\exp\lro{-r } r^{\nd-1}} {(s\abs{\widehat{h}}\sqrt{r \rd \nd })^{\tau\lro{\nd - 1}}}  I_{\nd-1}\bigl(2s\abs{\widehat{h}}\sqrt{r \rd \nd }\bigr)^{\tau}  \nonumber\\
&& \times\int_{0}^{\infty} \frac{\exp\lro{-\lro{\sigma\sub{p}^{-2} + \rd\nd}z}}{ \lro{\sqrt{r z \rd \nd}}^{\nd-1} }   I_{\nd-1}\lro{2\sqrt{r z \rd \nd}} I_0\bigl(2\abs{a(\widehat{h})} \sigma\sub{p}^{-2} \sqrt{z} \bigr) \mathrm{d}z \mathrm{d}r.\IEEEeqnarraynumspace
 \end{IEEEeqnarray}
Finally, we obtain~\eqref{eq:error_exponent_pat-nn} by using~\eqref{eq:f0_final} in~\eqref{eq:generalized_random_coding_error_exponent} and by taking an expectation over $\widehat{H}$. 

\subsection{Proof of Theorem \ref{thm:MC}} 
 \label{app:MC_proof}
We use as auxiliary channel in the min-max converse~\cite[Thm.~27]{polyanskiy10-05a}, the one for which $\vecy^\ell$ has \gls{pdf}
  \begin{IEEEeqnarray}{rCl}
  	Q_{\randvecy^\ell}\lro{\vecy^\ell} = \prod_{k=1}^{\ell} P_{\randvecy}\lro{\vecy_k}
  \end{IEEEeqnarray}
where $P_{\randvecy}$ is given in~\eqref{eq:shell_induced_out_pdf}.
Note now that for every $\nc\times \nc$ unitary matrix $\matV$, we have  $P_{\randvecy}\lro{\matV\vecy_k} = 
P_{\randvecy}\lro{\vecy_k}$ and $P_{\randvecy \given \randvecx}\lro{\vecy_k \given 
\herm{\matV}\vecx_k} = P_{\randvecy\given \randvecx}\lro{\matV \vecy_k \given \vecx_k}$.
Along with~\eqref{eq:Sk}, this imply that the Neyman-Pearson function $\beta\lro{\vecx^\ell, 
Q_{\randvecy^\ell}}$ defined in \cite[Eq. (105)]{polyanskiy10-05a} is independent of $\vecx^\ell$.
Hence, we can use~\cite[Thm. 28]{polyanskiy10-05a} to conclude that 
\Rmax is upper-bounded as
\begin{IEEEeqnarray}{rCl}
\label{eq:MC2}
	\Rmax \leq \frac{1}{\nc\ell} \log\frac{1}{\beta_{1-\epsilon} \lro{\vecx^\ell, Q_{\randvecy^\ell}}}.
\end{IEEEeqnarray}
Without loss of generality, we shall set $\vecx_k=[\sqrt{n_c\rho},0\dots,0]$, $k=1,\dots,\ell$.
It follows by the Neyman-Pearson lemma~\cite{neyman33-01a} that 
\begin{equation}
	\beta_{1-\epsilon} \lro{\vecx^\ell, q_{\randvecy^\ell}} = \Pr\lrbo{r^{\ell}\lro{\vecx^\ell, \randvecy^\ell} \geq \gamma}, \quad \randvecy^\ell \sim Q_{\randvecy^\ell} 
\end{equation}
where $\gamma$ is the solution to
\begin{equation}
	\Pr\lrbo{r^{\ell}\lro{\vecx^\ell, \randvecy^\ell} \leq \gamma} = \epsilon, \quad \randvecy^\ell \sim P_{\randvecy^\ell \given \randvecx^\ell}
\end{equation}
and
\begin{equation}
\label{eq:info_dens_miss}
r^{\ell}\lro{\vecx^\ell, \vecy^\ell} =\sum_{k=1}^\ell r\lro{\vecx_k, \vecy_k}=\sum_{k=1}^\ell \log\frac{P_{\randvecy \given \randvecx}\lro{\vecy_k \given \vecx_k}}{ P_{\randvecy}\lro{\vecy_k}}.
\end{equation}
Finally, we obtain~\eqref{eq:MC} by relaxing~\eqref{eq:MC2} using~\cite[Eq. (106)]{polyanskiy10-05a} (which yields a generalized Verd\'u-Han converse bound, cf.~\cite{han03-a}) and by exploiting that when $\randvecy_k \distas P_{\randvecy \given \randvecx=\vecx_k}$ the random variable $r\lro{\vecx_k, \randvecy_k}$ is distributed as $S_k^s$ in~\eqref{eq:Sk} with $s=1$.

\bibliographystyle{IEEEtran}
\bibliography{./Inputs/giubib}
\end{document}